\documentclass[useAMS,usenatbib]{mn2e}
\usepackage{times}
\usepackage{url}
\usepackage{graphicx}
\usepackage{deluxetable}
\usepackage[usenames]{color}
\DeclareGraphicsExtensions{.pdf,.png,.jpg,.mps,.eps,.ps}
\usepackage{amsmath}
\usepackage{natbib}
\usepackage{subfig}
\usepackage{soul}
\definecolor{AliceBlue}{rgb}{0.94,0.97,1.00}
\definecolor{AntiqueWhite1}{rgb}{1.00,0.94,0.86}
\definecolor{AntiqueWhite2}{rgb}{0.93,0.87,0.80}
\definecolor{AntiqueWhite3}{rgb}{0.80,0.75,0.69}
\definecolor{AntiqueWhite4}{rgb}{0.55,0.51,0.47}
\definecolor{AntiqueWhite}{rgb}{0.98,0.92,0.84}
\definecolor{BlanchedAlmond}{rgb}{1.00,0.92,0.80}
\definecolor{BlueViolet}{rgb}{0.54,0.17,0.89}
\definecolor{CadetBlue1}{rgb}{0.60,0.96,1.00}
\definecolor{CadetBlue2}{rgb}{0.56,0.90,0.93}
\definecolor{CadetBlue3}{rgb}{0.48,0.77,0.80}
\definecolor{CadetBlue4}{rgb}{0.33,0.53,0.55}
\definecolor{CadetBlue}{rgb}{0.37,0.62,0.63}
\definecolor{CornflowerBlue}{rgb}{0.39,0.58,0.93}
\definecolor{DarkBlue}{rgb}{0.00,0.00,0.55}
\definecolor{DarkCyan}{rgb}{0.00,0.55,0.55}
\definecolor{DarkGoldenrod1}{rgb}{1.00,0.73,0.06}
\definecolor{DarkGoldenrod2}{rgb}{0.93,0.68,0.05}
\definecolor{DarkGoldenrod3}{rgb}{0.80,0.58,0.05}
\definecolor{DarkGoldenrod4}{rgb}{0.55,0.40,0.03}
\definecolor{DarkGoldenrod}{rgb}{0.72,0.53,0.04}
\definecolor{DarkGray}{rgb}{0.66,0.66,0.66}
\definecolor{DarkGreen}{rgb}{0.00,0.39,0.00}
\definecolor{DarkGrey}{rgb}{0.66,0.66,0.66}
\definecolor{DarkKhaki}{rgb}{0.74,0.72,0.42}
\definecolor{DarkMagenta}{rgb}{0.55,0.00,0.55}
\definecolor{DarkOliveGreen1}{rgb}{0.79,1.00,0.44}
\definecolor{DarkOliveGreen2}{rgb}{0.74,0.93,0.41}
\definecolor{DarkOliveGreen3}{rgb}{0.64,0.80,0.35}
\definecolor{DarkOliveGreen4}{rgb}{0.43,0.55,0.24}
\definecolor{DarkOliveGreen}{rgb}{0.33,0.42,0.18}
\definecolor{DarkOrange1}{rgb}{1.00,0.50,0.00}
\definecolor{DarkOrange2}{rgb}{0.93,0.46,0.00}
\definecolor{DarkOrange3}{rgb}{0.80,0.40,0.00}
\definecolor{DarkOrange4}{rgb}{0.55,0.27,0.00}
\definecolor{DarkOrange}{rgb}{1.00,0.55,0.00}
\definecolor{DarkOrchid1}{rgb}{0.75,0.24,1.00}
\definecolor{DarkOrchid2}{rgb}{0.70,0.23,0.93}
\definecolor{DarkOrchid3}{rgb}{0.60,0.20,0.80}
\definecolor{DarkOrchid4}{rgb}{0.41,0.13,0.55}
\definecolor{DarkOrchid}{rgb}{0.60,0.20,0.80}
\definecolor{DarkRed}{rgb}{0.55,0.00,0.00}
\definecolor{DarkSalmon}{rgb}{0.91,0.59,0.48}
\definecolor{DarkSeaGreen1}{rgb}{0.76,1.00,0.76}
\definecolor{DarkSeaGreen2}{rgb}{0.71,0.93,0.71}
\definecolor{DarkSeaGreen3}{rgb}{0.61,0.80,0.61}
\definecolor{DarkSeaGreen4}{rgb}{0.41,0.55,0.41}
\definecolor{DarkSeaGreen}{rgb}{0.56,0.74,0.56}
\definecolor{DarkSlateBlue}{rgb}{0.28,0.24,0.55}
\definecolor{DarkSlateGray1}{rgb}{0.59,1.00,1.00}
\definecolor{DarkSlateGray2}{rgb}{0.55,0.93,0.93}
\definecolor{DarkSlateGray3}{rgb}{0.47,0.80,0.80}
\definecolor{DarkSlateGray4}{rgb}{0.32,0.55,0.55}
\definecolor{DarkSlateGray}{rgb}{0.18,0.31,0.31}
\definecolor{DarkSlateGrey}{rgb}{0.18,0.31,0.31}
\definecolor{DarkTurquoise}{rgb}{0.00,0.81,0.82}
\definecolor{DarkViolet}{rgb}{0.58,0.00,0.83}
\definecolor{DeepPink1}{rgb}{1.00,0.08,0.58}
\definecolor{DeepPink2}{rgb}{0.93,0.07,0.54}
\definecolor{DeepPink3}{rgb}{0.80,0.06,0.46}
\definecolor{DeepPink4}{rgb}{0.55,0.04,0.31}
\definecolor{DeepPink}{rgb}{1.00,0.08,0.58}
\definecolor{DeepSkyBlue1}{rgb}{0.00,0.75,1.00}
\definecolor{DeepSkyBlue2}{rgb}{0.00,0.70,0.93}
\definecolor{DeepSkyBlue3}{rgb}{0.00,0.60,0.80}
\definecolor{DeepSkyBlue4}{rgb}{0.00,0.41,0.55}
\definecolor{DeepSkyBlue}{rgb}{0.00,0.75,1.00}
\definecolor{DimGray}{rgb}{0.41,0.41,0.41}
\definecolor{DimGrey}{rgb}{0.41,0.41,0.41}
\definecolor{DodgerBlue1}{rgb}{0.12,0.56,1.00}
\definecolor{DodgerBlue2}{rgb}{0.11,0.53,0.93}
\definecolor{DodgerBlue3}{rgb}{0.09,0.45,0.80}
\definecolor{DodgerBlue4}{rgb}{0.06,0.31,0.55}
\definecolor{DodgerBlue}{rgb}{0.12,0.56,1.00}
\definecolor{FloralWhite}{rgb}{1.00,0.98,0.94}
\definecolor{ForestGreen}{rgb}{0.13,0.55,0.13}
\definecolor{GhostWhite}{rgb}{0.97,0.97,1.00}
\definecolor{GreenYellow}{rgb}{0.68,1.00,0.18}
\definecolor{HotPink1}{rgb}{1.00,0.43,0.71}
\definecolor{HotPink2}{rgb}{0.93,0.42,0.65}
\definecolor{HotPink3}{rgb}{0.80,0.38,0.56}
\definecolor{HotPink4}{rgb}{0.55,0.23,0.38}
\definecolor{HotPink}{rgb}{1.00,0.41,0.71}
\definecolor{IndianRed1}{rgb}{1.00,0.42,0.42}
\definecolor{IndianRed2}{rgb}{0.93,0.39,0.39}
\definecolor{IndianRed3}{rgb}{0.80,0.33,0.33}
\definecolor{IndianRed4}{rgb}{0.55,0.23,0.23}
\definecolor{IndianRed}{rgb}{0.80,0.36,0.36}
\definecolor{LavenderBlush1}{rgb}{1.00,0.94,0.96}
\definecolor{LavenderBlush2}{rgb}{0.93,0.88,0.90}
\definecolor{LavenderBlush3}{rgb}{0.80,0.76,0.77}
\definecolor{LavenderBlush4}{rgb}{0.55,0.51,0.53}
\definecolor{LavenderBlush}{rgb}{1.00,0.94,0.96}
\definecolor{LawnGreen}{rgb}{0.49,0.99,0.00}
\definecolor{LemonChiffon1}{rgb}{1.00,0.98,0.80}
\definecolor{LemonChiffon2}{rgb}{0.93,0.91,0.75}
\definecolor{LemonChiffon3}{rgb}{0.80,0.79,0.65}
\definecolor{LemonChiffon4}{rgb}{0.55,0.54,0.44}
\definecolor{LemonChiffon}{rgb}{1.00,0.98,0.80}
\definecolor{LightBlue1}{rgb}{0.75,0.94,1.00}
\definecolor{LightBlue2}{rgb}{0.70,0.87,0.93}
\definecolor{LightBlue3}{rgb}{0.60,0.75,0.80}
\definecolor{LightBlue4}{rgb}{0.41,0.51,0.55}
\definecolor{LightBlue}{rgb}{0.68,0.85,0.90}
\definecolor{LightCoral}{rgb}{0.94,0.50,0.50}
\definecolor{LightCyan1}{rgb}{0.88,1.00,1.00}
\definecolor{LightCyan2}{rgb}{0.82,0.93,0.93}
\definecolor{LightCyan3}{rgb}{0.71,0.80,0.80}
\definecolor{LightCyan4}{rgb}{0.48,0.55,0.55}
\definecolor{LightCyan}{rgb}{0.88,1.00,1.00}
\definecolor{LightGoldenrod1}{rgb}{1.00,0.93,0.55}
\definecolor{LightGoldenrod2}{rgb}{0.93,0.86,0.51}
\definecolor{LightGoldenrod3}{rgb}{0.80,0.75,0.44}
\definecolor{LightGoldenrod4}{rgb}{0.55,0.51,0.30}
\definecolor{LightGoldenrodYellow}{rgb}{0.98,0.98,0.82}
\definecolor{LightGoldenrod}{rgb}{0.93,0.87,0.51}
\definecolor{LightGray}{rgb}{0.83,0.83,0.83}
\definecolor{LightGreen}{rgb}{0.56,0.93,0.56}
\definecolor{LightGrey}{rgb}{0.83,0.83,0.83}
\definecolor{LightPink1}{rgb}{1.00,0.68,0.73}
\definecolor{LightPink2}{rgb}{0.93,0.64,0.68}
\definecolor{LightPink3}{rgb}{0.80,0.55,0.58}
\definecolor{LightPink4}{rgb}{0.55,0.37,0.40}
\definecolor{LightPink}{rgb}{1.00,0.71,0.76}
\definecolor{LightSalmon1}{rgb}{1.00,0.63,0.48}
\definecolor{LightSalmon2}{rgb}{0.93,0.58,0.45}
\definecolor{LightSalmon3}{rgb}{0.80,0.51,0.38}
\definecolor{LightSalmon4}{rgb}{0.55,0.34,0.26}
\definecolor{LightSalmon}{rgb}{1.00,0.63,0.48}
\definecolor{LightSeaGreen}{rgb}{0.13,0.70,0.67}
\definecolor{LightSkyBlue1}{rgb}{0.69,0.89,1.00}
\definecolor{LightSkyBlue2}{rgb}{0.64,0.83,0.93}
\definecolor{LightSkyBlue3}{rgb}{0.55,0.71,0.80}
\definecolor{LightSkyBlue4}{rgb}{0.38,0.48,0.55}
\definecolor{LightSkyBlue}{rgb}{0.53,0.81,0.98}
\definecolor{LightSlateBlue}{rgb}{0.52,0.44,1.00}
\definecolor{LightSlateGray}{rgb}{0.47,0.53,0.60}
\definecolor{LightSlateGrey}{rgb}{0.47,0.53,0.60}
\definecolor{LightSteelBlue1}{rgb}{0.79,0.88,1.00}
\definecolor{LightSteelBlue2}{rgb}{0.74,0.82,0.93}
\definecolor{LightSteelBlue3}{rgb}{0.64,0.71,0.80}
\definecolor{LightSteelBlue4}{rgb}{0.43,0.48,0.55}
\definecolor{LightSteelBlue}{rgb}{0.69,0.77,0.87}
\definecolor{LightYellow1}{rgb}{1.00,1.00,0.88}
\definecolor{LightYellow2}{rgb}{0.93,0.93,0.82}
\definecolor{LightYellow3}{rgb}{0.80,0.80,0.71}
\definecolor{LightYellow4}{rgb}{0.55,0.55,0.48}
\definecolor{LightYellow}{rgb}{1.00,1.00,0.88}
\definecolor{LimeGreen}{rgb}{0.20,0.80,0.20}
\definecolor{MediumAquamarine}{rgb}{0.40,0.80,0.67}
\definecolor{MediumBlue}{rgb}{0.00,0.00,0.80}
\definecolor{MediumOrchid1}{rgb}{0.88,0.40,1.00}
\definecolor{MediumOrchid2}{rgb}{0.82,0.37,0.93}
\definecolor{MediumOrchid3}{rgb}{0.71,0.32,0.80}
\definecolor{MediumOrchid4}{rgb}{0.48,0.22,0.55}
\definecolor{MediumOrchid}{rgb}{0.73,0.33,0.83}
\definecolor{MediumPurple1}{rgb}{0.67,0.51,1.00}
\definecolor{MediumPurple2}{rgb}{0.62,0.47,0.93}
\definecolor{MediumPurple3}{rgb}{0.54,0.41,0.80}
\definecolor{MediumPurple4}{rgb}{0.36,0.28,0.55}
\definecolor{MediumPurple}{rgb}{0.58,0.44,0.86}
\definecolor{MediumSeaGreen}{rgb}{0.24,0.70,0.44}
\definecolor{MediumSlateBlue}{rgb}{0.48,0.41,0.93}
\definecolor{MediumSpringGreen}{rgb}{0.00,0.98,0.60}
\definecolor{MediumTurquoise}{rgb}{0.28,0.82,0.80}
\definecolor{MediumVioletRed}{rgb}{0.78,0.08,0.52}
\definecolor{MidnightBlue}{rgb}{0.10,0.10,0.44}
\definecolor{MintCream}{rgb}{0.96,1.00,0.98}
\definecolor{MistyRose1}{rgb}{1.00,0.89,0.88}
\definecolor{MistyRose2}{rgb}{0.93,0.84,0.82}
\definecolor{MistyRose3}{rgb}{0.80,0.72,0.71}
\definecolor{MistyRose4}{rgb}{0.55,0.49,0.48}
\definecolor{MistyRose}{rgb}{1.00,0.89,0.88}
\definecolor{NavajoWhite1}{rgb}{1.00,0.87,0.68}
\definecolor{NavajoWhite2}{rgb}{0.93,0.81,0.63}
\definecolor{NavajoWhite3}{rgb}{0.80,0.70,0.55}
\definecolor{NavajoWhite4}{rgb}{0.55,0.47,0.37}
\definecolor{NavajoWhite}{rgb}{1.00,0.87,0.68}
\definecolor{NavyBlue}{rgb}{0.00,0.00,0.50}
\definecolor{OldLace}{rgb}{0.99,0.96,0.90}
\definecolor{OliveDrab1}{rgb}{0.75,1.00,0.24}
\definecolor{OliveDrab2}{rgb}{0.70,0.93,0.23}
\definecolor{OliveDrab3}{rgb}{0.60,0.80,0.20}
\definecolor{OliveDrab4}{rgb}{0.41,0.55,0.13}
\definecolor{OliveDrab}{rgb}{0.42,0.56,0.14}
\definecolor{OrangeRed1}{rgb}{1.00,0.27,0.00}
\definecolor{OrangeRed2}{rgb}{0.93,0.25,0.00}
\definecolor{OrangeRed3}{rgb}{0.80,0.22,0.00}
\definecolor{OrangeRed4}{rgb}{0.55,0.15,0.00}
\definecolor{OrangeRed}{rgb}{1.00,0.27,0.00}
\definecolor{PaleGoldenrod}{rgb}{0.93,0.91,0.67}
\definecolor{PaleGreen1}{rgb}{0.60,1.00,0.60}
\definecolor{PaleGreen2}{rgb}{0.56,0.93,0.56}
\definecolor{PaleGreen3}{rgb}{0.49,0.80,0.49}
\definecolor{PaleGreen4}{rgb}{0.33,0.55,0.33}
\definecolor{PaleGreen}{rgb}{0.60,0.98,0.60}
\definecolor{PaleTurquoise1}{rgb}{0.73,1.00,1.00}
\definecolor{PaleTurquoise2}{rgb}{0.68,0.93,0.93}
\definecolor{PaleTurquoise3}{rgb}{0.59,0.80,0.80}
\definecolor{PaleTurquoise4}{rgb}{0.40,0.55,0.55}
\definecolor{PaleTurquoise}{rgb}{0.69,0.93,0.93}
\definecolor{PaleVioletRed1}{rgb}{1.00,0.51,0.67}
\definecolor{PaleVioletRed2}{rgb}{0.93,0.47,0.62}
\definecolor{PaleVioletRed3}{rgb}{0.80,0.41,0.54}
\definecolor{PaleVioletRed4}{rgb}{0.55,0.28,0.36}
\definecolor{PaleVioletRed}{rgb}{0.86,0.44,0.58}
\definecolor{PapayaWhip}{rgb}{1.00,0.94,0.84}
\definecolor{PeachPuff1}{rgb}{1.00,0.85,0.73}
\definecolor{PeachPuff2}{rgb}{0.93,0.80,0.68}
\definecolor{PeachPuff3}{rgb}{0.80,0.69,0.58}
\definecolor{PeachPuff4}{rgb}{0.55,0.47,0.40}
\definecolor{PeachPuff}{rgb}{1.00,0.85,0.73}
\definecolor{PowderBlue}{rgb}{0.69,0.88,0.90}
\definecolor{RosyBrown1}{rgb}{1.00,0.76,0.76}
\definecolor{RosyBrown2}{rgb}{0.93,0.71,0.71}
\definecolor{RosyBrown3}{rgb}{0.80,0.61,0.61}
\definecolor{RosyBrown4}{rgb}{0.55,0.41,0.41}
\definecolor{RosyBrown}{rgb}{0.74,0.56,0.56}
\definecolor{RoyalBlue1}{rgb}{0.28,0.46,1.00}
\definecolor{RoyalBlue2}{rgb}{0.26,0.43,0.93}
\definecolor{RoyalBlue3}{rgb}{0.23,0.37,0.80}
\definecolor{RoyalBlue4}{rgb}{0.15,0.25,0.55}
\definecolor{RoyalBlue}{rgb}{0.25,0.41,0.88}
\definecolor{SaddleBrown}{rgb}{0.55,0.27,0.07}
\definecolor{SandyBrown}{rgb}{0.96,0.64,0.38}
\definecolor{SeaGreen1}{rgb}{0.33,1.00,0.62}
\definecolor{SeaGreen2}{rgb}{0.31,0.93,0.58}
\definecolor{SeaGreen3}{rgb}{0.26,0.80,0.50}
\definecolor{SeaGreen4}{rgb}{0.18,0.55,0.34}
\definecolor{SeaGreen}{rgb}{0.18,0.55,0.34}
\definecolor{SkyBlue1}{rgb}{0.53,0.81,1.00}
\definecolor{SkyBlue2}{rgb}{0.49,0.75,0.93}
\definecolor{SkyBlue3}{rgb}{0.42,0.65,0.80}
\definecolor{SkyBlue4}{rgb}{0.29,0.44,0.55}
\definecolor{SkyBlue}{rgb}{0.53,0.81,0.92}
\definecolor{SlateBlue1}{rgb}{0.51,0.44,1.00}
\definecolor{SlateBlue2}{rgb}{0.48,0.40,0.93}
\definecolor{SlateBlue3}{rgb}{0.41,0.35,0.80}
\definecolor{SlateBlue4}{rgb}{0.28,0.24,0.55}
\definecolor{SlateBlue}{rgb}{0.42,0.35,0.80}
\definecolor{SlateGray1}{rgb}{0.78,0.89,1.00}
\definecolor{SlateGray2}{rgb}{0.73,0.83,0.93}
\definecolor{SlateGray3}{rgb}{0.62,0.71,0.80}
\definecolor{SlateGray4}{rgb}{0.42,0.48,0.55}
\definecolor{SlateGray}{rgb}{0.44,0.50,0.56}
\definecolor{SlateGrey}{rgb}{0.44,0.50,0.56}
\definecolor{SpringGreen1}{rgb}{0.00,1.00,0.50}
\definecolor{SpringGreen2}{rgb}{0.00,0.93,0.46}
\definecolor{SpringGreen3}{rgb}{0.00,0.80,0.40}
\definecolor{SpringGreen4}{rgb}{0.00,0.55,0.27}
\definecolor{SpringGreen}{rgb}{0.00,1.00,0.50}
\definecolor{SteelBlue1}{rgb}{0.39,0.72,1.00}
\definecolor{SteelBlue2}{rgb}{0.36,0.67,0.93}
\definecolor{SteelBlue3}{rgb}{0.31,0.58,0.80}
\definecolor{SteelBlue4}{rgb}{0.21,0.39,0.55}
\definecolor{SteelBlue}{rgb}{0.27,0.51,0.71}
\definecolor{VioletRed1}{rgb}{1.00,0.24,0.59}
\definecolor{VioletRed2}{rgb}{0.93,0.23,0.55}
\definecolor{VioletRed3}{rgb}{0.80,0.20,0.47}
\definecolor{VioletRed4}{rgb}{0.55,0.13,0.32}
\definecolor{VioletRed}{rgb}{0.82,0.13,0.56}
\definecolor{WhiteSmoke}{rgb}{0.96,0.96,0.96}
\definecolor{YellowGreen}{rgb}{0.60,0.80,0.20}
\definecolor{aliceblue}{rgb}{0.94,0.97,1.00}
\definecolor{antiquewhite}{rgb}{0.98,0.92,0.84}
\definecolor{aquamarine1}{rgb}{0.50,1.00,0.83}
\definecolor{aquamarine2}{rgb}{0.46,0.93,0.78}
\definecolor{aquamarine3}{rgb}{0.40,0.80,0.67}
\definecolor{aquamarine4}{rgb}{0.27,0.55,0.45}
\definecolor{aquamarine}{rgb}{0.50,1.00,0.83}
\definecolor{azure1}{rgb}{0.94,1.00,1.00}
\definecolor{azure2}{rgb}{0.88,0.93,0.93}
\definecolor{azure3}{rgb}{0.76,0.80,0.80}
\definecolor{azure4}{rgb}{0.51,0.55,0.55}
\definecolor{azure}{rgb}{0.94,1.00,1.00}
\definecolor{beige}{rgb}{0.96,0.96,0.86}
\definecolor{bisque1}{rgb}{1.00,0.89,0.77}
\definecolor{bisque2}{rgb}{0.93,0.84,0.72}
\definecolor{bisque3}{rgb}{0.80,0.72,0.62}
\definecolor{bisque4}{rgb}{0.55,0.49,0.42}
\definecolor{bisque}{rgb}{1.00,0.89,0.77}
\definecolor{black}{rgb}{0.00,0.00,0.00}
\definecolor{blanchedalmond}{rgb}{1.00,0.92,0.80}
\definecolor{blue1}{rgb}{0.00,0.00,1.00}
\definecolor{blue2}{rgb}{0.00,0.00,0.93}
\definecolor{blue3}{rgb}{0.00,0.00,0.80}
\definecolor{blue4}{rgb}{0.00,0.00,0.55}
\definecolor{blueviolet}{rgb}{0.54,0.17,0.89}
\definecolor{blue}{rgb}{0.00,0.00,1.00}
\definecolor{brown1}{rgb}{1.00,0.25,0.25}
\definecolor{brown2}{rgb}{0.93,0.23,0.23}
\definecolor{brown3}{rgb}{0.80,0.20,0.20}
\definecolor{brown4}{rgb}{0.55,0.14,0.14}
\definecolor{brown}{rgb}{0.65,0.16,0.16}
\definecolor{burlywood1}{rgb}{1.00,0.83,0.61}
\definecolor{burlywood2}{rgb}{0.93,0.77,0.57}
\definecolor{burlywood3}{rgb}{0.80,0.67,0.49}
\definecolor{burlywood4}{rgb}{0.55,0.45,0.33}
\definecolor{burlywood}{rgb}{0.87,0.72,0.53}
\definecolor{cadetblue}{rgb}{0.37,0.62,0.63}
\definecolor{chartreuse1}{rgb}{0.50,1.00,0.00}
\definecolor{chartreuse2}{rgb}{0.46,0.93,0.00}
\definecolor{chartreuse3}{rgb}{0.40,0.80,0.00}
\definecolor{chartreuse4}{rgb}{0.27,0.55,0.00}
\definecolor{chartreuse}{rgb}{0.50,1.00,0.00}
\definecolor{chocolate1}{rgb}{1.00,0.50,0.14}
\definecolor{chocolate2}{rgb}{0.93,0.46,0.13}
\definecolor{chocolate3}{rgb}{0.80,0.40,0.11}
\definecolor{chocolate4}{rgb}{0.55,0.27,0.07}
\definecolor{chocolate}{rgb}{0.82,0.41,0.12}
\definecolor{coral1}{rgb}{1.00,0.45,0.34}
\definecolor{coral2}{rgb}{0.93,0.42,0.31}
\definecolor{coral3}{rgb}{0.80,0.36,0.27}
\definecolor{coral4}{rgb}{0.55,0.24,0.18}
\definecolor{coral}{rgb}{1.00,0.50,0.31}
\definecolor{cornflowerblue}{rgb}{0.39,0.58,0.93}
\definecolor{cornsilk1}{rgb}{1.00,0.97,0.86}
\definecolor{cornsilk2}{rgb}{0.93,0.91,0.80}
\definecolor{cornsilk3}{rgb}{0.80,0.78,0.69}
\definecolor{cornsilk4}{rgb}{0.55,0.53,0.47}
\definecolor{cornsilk}{rgb}{1.00,0.97,0.86}
\definecolor{cyan1}{rgb}{0.00,1.00,1.00}
\definecolor{cyan2}{rgb}{0.00,0.93,0.93}
\definecolor{cyan3}{rgb}{0.00,0.80,0.80}
\definecolor{cyan4}{rgb}{0.00,0.55,0.55}
\definecolor{cyan}{rgb}{0.00,1.00,1.00}
\definecolor{darkblue}{rgb}{0.00,0.00,0.55}
\definecolor{darkcyan}{rgb}{0.00,0.55,0.55}
\definecolor{darkgoldenrod}{rgb}{0.72,0.53,0.04}
\definecolor{darkgray}{rgb}{0.66,0.66,0.66}
\definecolor{darkgreen}{rgb}{0.00,0.39,0.00}
\definecolor{darkgrey}{rgb}{0.66,0.66,0.66}
\definecolor{darkkhaki}{rgb}{0.74,0.72,0.42}
\definecolor{darkmagenta}{rgb}{0.55,0.00,0.55}
\definecolor{darkolive}{rgb}{0.33,0.42,0.18}
\definecolor{darkorange}{rgb}{1.00,0.55,0.00}
\definecolor{darkorchid}{rgb}{0.60,0.20,0.80}
\definecolor{darkred}{rgb}{0.55,0.00,0.00}
\definecolor{darksalmon}{rgb}{0.91,0.59,0.48}
\definecolor{darksea}{rgb}{0.56,0.74,0.56}
\definecolor{darkslate}{rgb}{0.18,0.31,0.31}
\definecolor{darkslate}{rgb}{0.18,0.31,0.31}
\definecolor{darkslate}{rgb}{0.28,0.24,0.55}
\definecolor{darkturquoise}{rgb}{0.00,0.81,0.82}
\definecolor{darkviolet}{rgb}{0.58,0.00,0.83}
\definecolor{deeppink}{rgb}{1.00,0.08,0.58}
\definecolor{deepsky}{rgb}{0.00,0.75,1.00}
\definecolor{dimgray}{rgb}{0.41,0.41,0.41}
\definecolor{dimgrey}{rgb}{0.41,0.41,0.41}
\definecolor{dodgerblue}{rgb}{0.12,0.56,1.00}
\definecolor{firebrick1}{rgb}{1.00,0.19,0.19}
\definecolor{firebrick2}{rgb}{0.93,0.17,0.17}
\definecolor{firebrick3}{rgb}{0.80,0.15,0.15}
\definecolor{firebrick4}{rgb}{0.55,0.10,0.10}
\definecolor{firebrick}{rgb}{0.70,0.13,0.13}
\definecolor{floralwhite}{rgb}{1.00,0.98,0.94}
\definecolor{forestgreen}{rgb}{0.13,0.55,0.13}
\definecolor{gainsboro}{rgb}{0.86,0.86,0.86}
\definecolor{ghostwhite}{rgb}{0.97,0.97,1.00}
\definecolor{gold1}{rgb}{1.00,0.84,0.00}
\definecolor{gold2}{rgb}{0.93,0.79,0.00}
\definecolor{gold3}{rgb}{0.80,0.68,0.00}
\definecolor{gold4}{rgb}{0.55,0.46,0.00}
\definecolor{goldenrod1}{rgb}{1.00,0.76,0.15}
\definecolor{goldenrod2}{rgb}{0.93,0.71,0.13}
\definecolor{goldenrod3}{rgb}{0.80,0.61,0.11}
\definecolor{goldenrod4}{rgb}{0.55,0.41,0.08}
\definecolor{goldenrod}{rgb}{0.85,0.65,0.13}
\definecolor{gold}{rgb}{1.00,0.84,0.00}
\definecolor{gray0}{rgb}{0.00,0.00,0.00}
\definecolor{gray100}{rgb}{1.00,1.00,1.00}
\definecolor{gray10}{rgb}{0.10,0.10,0.10}
\definecolor{gray11}{rgb}{0.11,0.11,0.11}
\definecolor{gray12}{rgb}{0.12,0.12,0.12}
\definecolor{gray13}{rgb}{0.13,0.13,0.13}
\definecolor{gray14}{rgb}{0.14,0.14,0.14}
\definecolor{gray15}{rgb}{0.15,0.15,0.15}
\definecolor{gray16}{rgb}{0.16,0.16,0.16}
\definecolor{gray17}{rgb}{0.17,0.17,0.17}
\definecolor{gray18}{rgb}{0.18,0.18,0.18}
\definecolor{gray19}{rgb}{0.19,0.19,0.19}
\definecolor{gray1}{rgb}{0.01,0.01,0.01}
\definecolor{gray20}{rgb}{0.20,0.20,0.20}
\definecolor{gray21}{rgb}{0.21,0.21,0.21}
\definecolor{gray22}{rgb}{0.22,0.22,0.22}
\definecolor{gray23}{rgb}{0.23,0.23,0.23}
\definecolor{gray24}{rgb}{0.24,0.24,0.24}
\definecolor{gray25}{rgb}{0.25,0.25,0.25}
\definecolor{gray26}{rgb}{0.26,0.26,0.26}
\definecolor{gray27}{rgb}{0.27,0.27,0.27}
\definecolor{gray28}{rgb}{0.28,0.28,0.28}
\definecolor{gray29}{rgb}{0.29,0.29,0.29}
\definecolor{gray2}{rgb}{0.02,0.02,0.02}
\definecolor{gray30}{rgb}{0.30,0.30,0.30}
\definecolor{gray31}{rgb}{0.31,0.31,0.31}
\definecolor{gray32}{rgb}{0.32,0.32,0.32}
\definecolor{gray33}{rgb}{0.33,0.33,0.33}
\definecolor{gray34}{rgb}{0.34,0.34,0.34}
\definecolor{gray35}{rgb}{0.35,0.35,0.35}
\definecolor{gray36}{rgb}{0.36,0.36,0.36}
\definecolor{gray37}{rgb}{0.37,0.37,0.37}
\definecolor{gray38}{rgb}{0.38,0.38,0.38}
\definecolor{gray39}{rgb}{0.39,0.39,0.39}
\definecolor{gray3}{rgb}{0.03,0.03,0.03}
\definecolor{gray40}{rgb}{0.40,0.40,0.40}
\definecolor{gray41}{rgb}{0.41,0.41,0.41}
\definecolor{gray42}{rgb}{0.42,0.42,0.42}
\definecolor{gray43}{rgb}{0.43,0.43,0.43}
\definecolor{gray44}{rgb}{0.44,0.44,0.44}
\definecolor{gray45}{rgb}{0.45,0.45,0.45}
\definecolor{gray46}{rgb}{0.46,0.46,0.46}
\definecolor{gray47}{rgb}{0.47,0.47,0.47}
\definecolor{gray48}{rgb}{0.48,0.48,0.48}
\definecolor{gray49}{rgb}{0.49,0.49,0.49}
\definecolor{gray4}{rgb}{0.04,0.04,0.04}
\definecolor{gray50}{rgb}{0.50,0.50,0.50}
\definecolor{gray51}{rgb}{0.51,0.51,0.51}
\definecolor{gray52}{rgb}{0.52,0.52,0.52}
\definecolor{gray53}{rgb}{0.53,0.53,0.53}
\definecolor{gray54}{rgb}{0.54,0.54,0.54}
\definecolor{gray55}{rgb}{0.55,0.55,0.55}
\definecolor{gray56}{rgb}{0.56,0.56,0.56}
\definecolor{gray57}{rgb}{0.57,0.57,0.57}
\definecolor{gray58}{rgb}{0.58,0.58,0.58}
\definecolor{gray59}{rgb}{0.59,0.59,0.59}
\definecolor{gray5}{rgb}{0.05,0.05,0.05}
\definecolor{gray60}{rgb}{0.60,0.60,0.60}
\definecolor{gray61}{rgb}{0.61,0.61,0.61}
\definecolor{gray62}{rgb}{0.62,0.62,0.62}
\definecolor{gray63}{rgb}{0.63,0.63,0.63}
\definecolor{gray64}{rgb}{0.64,0.64,0.64}
\definecolor{gray65}{rgb}{0.65,0.65,0.65}
\definecolor{gray66}{rgb}{0.66,0.66,0.66}
\definecolor{gray67}{rgb}{0.67,0.67,0.67}
\definecolor{gray68}{rgb}{0.68,0.68,0.68}
\definecolor{gray69}{rgb}{0.69,0.69,0.69}
\definecolor{gray6}{rgb}{0.06,0.06,0.06}
\definecolor{gray70}{rgb}{0.70,0.70,0.70}
\definecolor{gray71}{rgb}{0.71,0.71,0.71}
\definecolor{gray72}{rgb}{0.72,0.72,0.72}
\definecolor{gray73}{rgb}{0.73,0.73,0.73}
\definecolor{gray74}{rgb}{0.74,0.74,0.74}
\definecolor{gray75}{rgb}{0.75,0.75,0.75}
\definecolor{gray76}{rgb}{0.76,0.76,0.76}
\definecolor{gray77}{rgb}{0.77,0.77,0.77}
\definecolor{gray78}{rgb}{0.78,0.78,0.78}
\definecolor{gray79}{rgb}{0.79,0.79,0.79}
\definecolor{gray7}{rgb}{0.07,0.07,0.07}
\definecolor{gray80}{rgb}{0.80,0.80,0.80}
\definecolor{gray81}{rgb}{0.81,0.81,0.81}
\definecolor{gray82}{rgb}{0.82,0.82,0.82}
\definecolor{gray83}{rgb}{0.83,0.83,0.83}
\definecolor{gray84}{rgb}{0.84,0.84,0.84}
\definecolor{gray85}{rgb}{0.85,0.85,0.85}
\definecolor{gray86}{rgb}{0.86,0.86,0.86}
\definecolor{gray87}{rgb}{0.87,0.87,0.87}
\definecolor{gray88}{rgb}{0.88,0.88,0.88}
\definecolor{gray89}{rgb}{0.89,0.89,0.89}
\definecolor{gray8}{rgb}{0.08,0.08,0.08}
\definecolor{gray90}{rgb}{0.90,0.90,0.90}
\definecolor{gray91}{rgb}{0.91,0.91,0.91}
\definecolor{gray92}{rgb}{0.92,0.92,0.92}
\definecolor{gray93}{rgb}{0.93,0.93,0.93}
\definecolor{gray94}{rgb}{0.94,0.94,0.94}
\definecolor{gray95}{rgb}{0.95,0.95,0.95}
\definecolor{gray96}{rgb}{0.96,0.96,0.96}
\definecolor{gray97}{rgb}{0.97,0.97,0.97}
\definecolor{gray98}{rgb}{0.98,0.98,0.98}
\definecolor{gray99}{rgb}{0.99,0.99,0.99}
\definecolor{gray9}{rgb}{0.09,0.09,0.09}
\definecolor{gray}{rgb}{0.75,0.75,0.75}
\definecolor{green1}{rgb}{0.00,1.00,0.00}
\definecolor{green2}{rgb}{0.00,0.93,0.00}
\definecolor{green3}{rgb}{0.00,0.80,0.00}
\definecolor{green4}{rgb}{0.00,0.55,0.00}
\definecolor{greenyellow}{rgb}{0.68,1.00,0.18}
\definecolor{green}{rgb}{0.00,1.00,0.00}
\definecolor{grey0}{rgb}{0.00,0.00,0.00}
\definecolor{grey100}{rgb}{1.00,1.00,1.00}
\definecolor{grey10}{rgb}{0.10,0.10,0.10}
\definecolor{grey11}{rgb}{0.11,0.11,0.11}
\definecolor{grey12}{rgb}{0.12,0.12,0.12}
\definecolor{grey13}{rgb}{0.13,0.13,0.13}
\definecolor{grey14}{rgb}{0.14,0.14,0.14}
\definecolor{grey15}{rgb}{0.15,0.15,0.15}
\definecolor{grey16}{rgb}{0.16,0.16,0.16}
\definecolor{grey17}{rgb}{0.17,0.17,0.17}
\definecolor{grey18}{rgb}{0.18,0.18,0.18}
\definecolor{grey19}{rgb}{0.19,0.19,0.19}
\definecolor{grey1}{rgb}{0.01,0.01,0.01}
\definecolor{grey20}{rgb}{0.20,0.20,0.20}
\definecolor{grey21}{rgb}{0.21,0.21,0.21}
\definecolor{grey22}{rgb}{0.22,0.22,0.22}
\definecolor{grey23}{rgb}{0.23,0.23,0.23}
\definecolor{grey24}{rgb}{0.24,0.24,0.24}
\definecolor{grey25}{rgb}{0.25,0.25,0.25}
\definecolor{grey26}{rgb}{0.26,0.26,0.26}
\definecolor{grey27}{rgb}{0.27,0.27,0.27}
\definecolor{grey28}{rgb}{0.28,0.28,0.28}
\definecolor{grey29}{rgb}{0.29,0.29,0.29}
\definecolor{grey2}{rgb}{0.02,0.02,0.02}
\definecolor{grey30}{rgb}{0.30,0.30,0.30}
\definecolor{grey31}{rgb}{0.31,0.31,0.31}
\definecolor{grey32}{rgb}{0.32,0.32,0.32}
\definecolor{grey33}{rgb}{0.33,0.33,0.33}
\definecolor{grey34}{rgb}{0.34,0.34,0.34}
\definecolor{grey35}{rgb}{0.35,0.35,0.35}
\definecolor{grey36}{rgb}{0.36,0.36,0.36}
\definecolor{grey37}{rgb}{0.37,0.37,0.37}
\definecolor{grey38}{rgb}{0.38,0.38,0.38}
\definecolor{grey39}{rgb}{0.39,0.39,0.39}
\definecolor{grey3}{rgb}{0.03,0.03,0.03}
\definecolor{grey40}{rgb}{0.40,0.40,0.40}
\definecolor{grey41}{rgb}{0.41,0.41,0.41}
\definecolor{grey42}{rgb}{0.42,0.42,0.42}
\definecolor{grey43}{rgb}{0.43,0.43,0.43}
\definecolor{grey44}{rgb}{0.44,0.44,0.44}
\definecolor{grey45}{rgb}{0.45,0.45,0.45}
\definecolor{grey46}{rgb}{0.46,0.46,0.46}
\definecolor{grey47}{rgb}{0.47,0.47,0.47}
\definecolor{grey48}{rgb}{0.48,0.48,0.48}
\definecolor{grey49}{rgb}{0.49,0.49,0.49}
\definecolor{grey4}{rgb}{0.04,0.04,0.04}
\definecolor{grey50}{rgb}{0.50,0.50,0.50}
\definecolor{grey51}{rgb}{0.51,0.51,0.51}
\definecolor{grey52}{rgb}{0.52,0.52,0.52}
\definecolor{grey53}{rgb}{0.53,0.53,0.53}
\definecolor{grey54}{rgb}{0.54,0.54,0.54}
\definecolor{grey55}{rgb}{0.55,0.55,0.55}
\definecolor{grey56}{rgb}{0.56,0.56,0.56}
\definecolor{grey57}{rgb}{0.57,0.57,0.57}
\definecolor{grey58}{rgb}{0.58,0.58,0.58}
\definecolor{grey59}{rgb}{0.59,0.59,0.59}
\definecolor{grey5}{rgb}{0.05,0.05,0.05}
\definecolor{grey60}{rgb}{0.60,0.60,0.60}
\definecolor{grey61}{rgb}{0.61,0.61,0.61}
\definecolor{grey62}{rgb}{0.62,0.62,0.62}
\definecolor{grey63}{rgb}{0.63,0.63,0.63}
\definecolor{grey64}{rgb}{0.64,0.64,0.64}
\definecolor{grey65}{rgb}{0.65,0.65,0.65}
\definecolor{grey66}{rgb}{0.66,0.66,0.66}
\definecolor{grey67}{rgb}{0.67,0.67,0.67}
\definecolor{grey68}{rgb}{0.68,0.68,0.68}
\definecolor{grey69}{rgb}{0.69,0.69,0.69}
\definecolor{grey6}{rgb}{0.06,0.06,0.06}
\definecolor{grey70}{rgb}{0.70,0.70,0.70}
\definecolor{grey71}{rgb}{0.71,0.71,0.71}
\definecolor{grey72}{rgb}{0.72,0.72,0.72}
\definecolor{grey73}{rgb}{0.73,0.73,0.73}
\definecolor{grey74}{rgb}{0.74,0.74,0.74}
\definecolor{grey75}{rgb}{0.75,0.75,0.75}
\definecolor{grey76}{rgb}{0.76,0.76,0.76}
\definecolor{grey77}{rgb}{0.77,0.77,0.77}
\definecolor{grey78}{rgb}{0.78,0.78,0.78}
\definecolor{grey79}{rgb}{0.79,0.79,0.79}
\definecolor{grey7}{rgb}{0.07,0.07,0.07}
\definecolor{grey80}{rgb}{0.80,0.80,0.80}
\definecolor{grey81}{rgb}{0.81,0.81,0.81}
\definecolor{grey82}{rgb}{0.82,0.82,0.82}
\definecolor{grey83}{rgb}{0.83,0.83,0.83}
\definecolor{grey84}{rgb}{0.84,0.84,0.84}
\definecolor{grey85}{rgb}{0.85,0.85,0.85}
\definecolor{grey86}{rgb}{0.86,0.86,0.86}
\definecolor{grey87}{rgb}{0.87,0.87,0.87}
\definecolor{grey88}{rgb}{0.88,0.88,0.88}
\definecolor{grey89}{rgb}{0.89,0.89,0.89}
\definecolor{grey8}{rgb}{0.08,0.08,0.08}
\definecolor{grey90}{rgb}{0.90,0.90,0.90}
\definecolor{grey91}{rgb}{0.91,0.91,0.91}
\definecolor{grey92}{rgb}{0.92,0.92,0.92}
\definecolor{grey93}{rgb}{0.93,0.93,0.93}
\definecolor{grey94}{rgb}{0.94,0.94,0.94}
\definecolor{grey95}{rgb}{0.95,0.95,0.95}
\definecolor{grey96}{rgb}{0.96,0.96,0.96}
\definecolor{grey97}{rgb}{0.97,0.97,0.97}
\definecolor{grey98}{rgb}{0.98,0.98,0.98}
\definecolor{grey99}{rgb}{0.99,0.99,0.99}
\definecolor{grey9}{rgb}{0.09,0.09,0.09}
\definecolor{grey}{rgb}{0.75,0.75,0.75}
\definecolor{honeydew1}{rgb}{0.94,1.00,0.94}
\definecolor{honeydew2}{rgb}{0.88,0.93,0.88}
\definecolor{honeydew3}{rgb}{0.76,0.80,0.76}
\definecolor{honeydew4}{rgb}{0.51,0.55,0.51}
\definecolor{honeydew}{rgb}{0.94,1.00,0.94}
\definecolor{hotpink}{rgb}{1.00,0.41,0.71}
\definecolor{indianred}{rgb}{0.80,0.36,0.36}
\definecolor{ivory1}{rgb}{1.00,1.00,0.94}
\definecolor{ivory2}{rgb}{0.93,0.93,0.88}
\definecolor{ivory3}{rgb}{0.80,0.80,0.76}
\definecolor{ivory4}{rgb}{0.55,0.55,0.51}
\definecolor{ivory}{rgb}{1.00,1.00,0.94}
\definecolor{khaki1}{rgb}{1.00,0.96,0.56}
\definecolor{khaki2}{rgb}{0.93,0.90,0.52}
\definecolor{khaki3}{rgb}{0.80,0.78,0.45}
\definecolor{khaki4}{rgb}{0.55,0.53,0.31}
\definecolor{khaki}{rgb}{0.94,0.90,0.55}
\definecolor{lavenderblush}{rgb}{1.00,0.94,0.96}
\definecolor{lavender}{rgb}{0.90,0.90,0.98}
\definecolor{lawngreen}{rgb}{0.49,0.99,0.00}
\definecolor{lemonchiffon}{rgb}{1.00,0.98,0.80}
\definecolor{lightblue}{rgb}{0.68,0.85,0.90}
\definecolor{lightcoral}{rgb}{0.94,0.50,0.50}
\definecolor{lightcyan}{rgb}{0.88,1.00,1.00}
\definecolor{lightgoldenrod}{rgb}{0.93,0.87,0.51}
\definecolor{lightgoldenrod}{rgb}{0.98,0.98,0.82}
\definecolor{lightgray}{rgb}{0.83,0.83,0.83}
\definecolor{lightgreen}{rgb}{0.56,0.93,0.56}
\definecolor{lightgrey}{rgb}{0.83,0.83,0.83}
\definecolor{lightpink}{rgb}{1.00,0.71,0.76}
\definecolor{lightsalmon}{rgb}{1.00,0.63,0.48}
\definecolor{lightsea}{rgb}{0.13,0.70,0.67}
\definecolor{lightsky}{rgb}{0.53,0.81,0.98}
\definecolor{lightslate}{rgb}{0.47,0.53,0.60}
\definecolor{lightslate}{rgb}{0.47,0.53,0.60}
\definecolor{lightslate}{rgb}{0.52,0.44,1.00}
\definecolor{lightsteel}{rgb}{0.69,0.77,0.87}
\definecolor{lightyellow}{rgb}{1.00,1.00,0.88}
\definecolor{limegreen}{rgb}{0.20,0.80,0.20}
\definecolor{linen}{rgb}{0.98,0.94,0.90}
\definecolor{magenta1}{rgb}{1.00,0.00,1.00}
\definecolor{magenta2}{rgb}{0.93,0.00,0.93}
\definecolor{magenta3}{rgb}{0.80,0.00,0.80}
\definecolor{magenta4}{rgb}{0.55,0.00,0.55}
\definecolor{magenta}{rgb}{1.00,0.00,1.00}
\definecolor{maroon1}{rgb}{1.00,0.20,0.70}
\definecolor{maroon2}{rgb}{0.93,0.19,0.65}
\definecolor{maroon3}{rgb}{0.80,0.16,0.56}
\definecolor{maroon4}{rgb}{0.55,0.11,0.38}
\definecolor{maroon}{rgb}{0.69,0.19,0.38}
\definecolor{mediumaquamarine}{rgb}{0.40,0.80,0.67}
\definecolor{mediumblue}{rgb}{0.00,0.00,0.80}
\definecolor{mediumorchid}{rgb}{0.73,0.33,0.83}
\definecolor{mediumpurple}{rgb}{0.58,0.44,0.86}
\definecolor{mediumsea}{rgb}{0.24,0.70,0.44}
\definecolor{mediumslate}{rgb}{0.48,0.41,0.93}
\definecolor{mediumspring}{rgb}{0.00,0.98,0.60}
\definecolor{mediumturquoise}{rgb}{0.28,0.82,0.80}
\definecolor{mediumviolet}{rgb}{0.78,0.08,0.52}
\definecolor{midnightblue}{rgb}{0.10,0.10,0.44}
\definecolor{mintcream}{rgb}{0.96,1.00,0.98}
\definecolor{mistyrose}{rgb}{1.00,0.89,0.88}
\definecolor{moccasin}{rgb}{1.00,0.89,0.71}
\definecolor{navajowhite}{rgb}{1.00,0.87,0.68}
\definecolor{navyblue}{rgb}{0.00,0.00,0.50}
\definecolor{navy}{rgb}{0.00,0.00,0.50}
\definecolor{oldlace}{rgb}{0.99,0.96,0.90}
\definecolor{olivedrab}{rgb}{0.42,0.56,0.14}
\definecolor{orange1}{rgb}{1.00,0.65,0.00}
\definecolor{orange2}{rgb}{0.93,0.60,0.00}
\definecolor{orange3}{rgb}{0.80,0.52,0.00}
\definecolor{orange4}{rgb}{0.55,0.35,0.00}
\definecolor{orangered}{rgb}{1.00,0.27,0.00}
\definecolor{orange}{rgb}{1.00,0.65,0.00}
\definecolor{orchid1}{rgb}{1.00,0.51,0.98}
\definecolor{orchid2}{rgb}{0.93,0.48,0.91}
\definecolor{orchid3}{rgb}{0.80,0.41,0.79}
\definecolor{orchid4}{rgb}{0.55,0.28,0.54}
\definecolor{orchid}{rgb}{0.85,0.44,0.84}
\definecolor{palegoldenrod}{rgb}{0.93,0.91,0.67}
\definecolor{palegreen}{rgb}{0.60,0.98,0.60}
\definecolor{paleturquoise}{rgb}{0.69,0.93,0.93}
\definecolor{paleviolet}{rgb}{0.86,0.44,0.58}
\definecolor{papayawhip}{rgb}{1.00,0.94,0.84}
\definecolor{peachpuff}{rgb}{1.00,0.85,0.73}
\definecolor{peru}{rgb}{0.80,0.52,0.25}
\definecolor{pink1}{rgb}{1.00,0.71,0.77}
\definecolor{pink2}{rgb}{0.93,0.66,0.72}
\definecolor{pink3}{rgb}{0.80,0.57,0.62}
\definecolor{pink4}{rgb}{0.55,0.39,0.42}
\definecolor{pink}{rgb}{1.00,0.75,0.80}
\definecolor{plum1}{rgb}{1.00,0.73,1.00}
\definecolor{plum2}{rgb}{0.93,0.68,0.93}
\definecolor{plum3}{rgb}{0.80,0.59,0.80}
\definecolor{plum4}{rgb}{0.55,0.40,0.55}
\definecolor{plum}{rgb}{0.87,0.63,0.87}
\definecolor{powderblue}{rgb}{0.69,0.88,0.90}
\definecolor{purple1}{rgb}{0.61,0.19,1.00}
\definecolor{purple2}{rgb}{0.57,0.17,0.93}
\definecolor{purple3}{rgb}{0.49,0.15,0.80}
\definecolor{purple4}{rgb}{0.33,0.10,0.55}
\definecolor{purple}{rgb}{0.63,0.13,0.94}
\definecolor{red1}{rgb}{1.00,0.00,0.00}
\definecolor{red2}{rgb}{0.93,0.00,0.00}
\definecolor{red3}{rgb}{0.80,0.00,0.00}
\definecolor{red4}{rgb}{0.55,0.00,0.00}
\definecolor{red}{rgb}{1.00,0.00,0.00}
\definecolor{rosybrown}{rgb}{0.74,0.56,0.56}
\definecolor{royalblue}{rgb}{0.25,0.41,0.88}
\definecolor{saddlebrown}{rgb}{0.55,0.27,0.07}
\definecolor{salmon1}{rgb}{1.00,0.55,0.41}
\definecolor{salmon2}{rgb}{0.93,0.51,0.38}
\definecolor{salmon3}{rgb}{0.80,0.44,0.33}
\definecolor{salmon4}{rgb}{0.55,0.30,0.22}
\definecolor{salmon}{rgb}{0.98,0.50,0.45}
\definecolor{sandybrown}{rgb}{0.96,0.64,0.38}
\definecolor{seagreen}{rgb}{0.18,0.55,0.34}
\definecolor{seashell1}{rgb}{1.00,0.96,0.93}
\definecolor{seashell2}{rgb}{0.93,0.90,0.87}
\definecolor{seashell3}{rgb}{0.80,0.77,0.75}
\definecolor{seashell4}{rgb}{0.55,0.53,0.51}
\definecolor{seashell}{rgb}{1.00,0.96,0.93}
\definecolor{sienna1}{rgb}{1.00,0.51,0.28}
\definecolor{sienna2}{rgb}{0.93,0.47,0.26}
\definecolor{sienna3}{rgb}{0.80,0.41,0.22}
\definecolor{sienna4}{rgb}{0.55,0.28,0.15}
\definecolor{sienna}{rgb}{0.63,0.32,0.18}
\definecolor{skyblue}{rgb}{0.53,0.81,0.92}
\definecolor{slateblue}{rgb}{0.42,0.35,0.80}
\definecolor{slategray}{rgb}{0.44,0.50,0.56}
\definecolor{slategrey}{rgb}{0.44,0.50,0.56}
\definecolor{snow1}{rgb}{1.00,0.98,0.98}
\definecolor{snow2}{rgb}{0.93,0.91,0.91}
\definecolor{snow3}{rgb}{0.80,0.79,0.79}
\definecolor{snow4}{rgb}{0.55,0.54,0.54}
\definecolor{snow}{rgb}{1.00,0.98,0.98}
\definecolor{springgreen}{rgb}{0.00,1.00,0.50}
\definecolor{steelblue}{rgb}{0.27,0.51,0.71}
\definecolor{tan1}{rgb}{1.00,0.65,0.31}
\definecolor{tan2}{rgb}{0.93,0.60,0.29}
\definecolor{tan3}{rgb}{0.80,0.52,0.25}
\definecolor{tan4}{rgb}{0.55,0.35,0.17}
\definecolor{tan}{rgb}{0.82,0.71,0.55}
\definecolor{thistle1}{rgb}{1.00,0.88,1.00}
\definecolor{thistle2}{rgb}{0.93,0.82,0.93}
\definecolor{thistle3}{rgb}{0.80,0.71,0.80}
\definecolor{thistle4}{rgb}{0.55,0.48,0.55}
\definecolor{thistle}{rgb}{0.85,0.75,0.85}
\definecolor{tomato1}{rgb}{1.00,0.39,0.28}
\definecolor{tomato2}{rgb}{0.93,0.36,0.26}
\definecolor{tomato3}{rgb}{0.80,0.31,0.22}
\definecolor{tomato4}{rgb}{0.55,0.21,0.15}
\definecolor{tomato}{rgb}{1.00,0.39,0.28}
\definecolor{turquoise1}{rgb}{0.00,0.96,1.00}
\definecolor{turquoise2}{rgb}{0.00,0.90,0.93}
\definecolor{turquoise3}{rgb}{0.00,0.77,0.80}
\definecolor{turquoise4}{rgb}{0.00,0.53,0.55}
\definecolor{turquoise}{rgb}{0.25,0.88,0.82}
\definecolor{violetred}{rgb}{0.82,0.13,0.56}
\definecolor{violet}{rgb}{0.93,0.51,0.93}
\definecolor{wheat1}{rgb}{1.00,0.91,0.73}
\definecolor{wheat2}{rgb}{0.93,0.85,0.68}
\definecolor{wheat3}{rgb}{0.80,0.73,0.59}
\definecolor{wheat4}{rgb}{0.55,0.49,0.40}
\definecolor{wheat}{rgb}{0.96,0.87,0.70}
\definecolor{whitesmoke}{rgb}{0.96,0.96,0.96}
\definecolor{white}{rgb}{1.00,1.00,1.00}
\definecolor{yellow1}{rgb}{1.00,1.00,0.00}
\definecolor{yellow2}{rgb}{0.93,0.93,0.00}
\definecolor{yellow3}{rgb}{0.80,0.80,0.00}
\definecolor{yellow4}{rgb}{0.55,0.55,0.00}
\definecolor{yellowgreen}{rgb}{0.60,0.80,0.20}
\definecolor{yellow}{rgb}{1.00,1.00,0.00}

\usepackage{lscape}
\usepackage{longtable}

\newcommand\swift{{\it Swift}}

\newcommand\rxte{{\it RXTE}}

\newcommand\astrosat{{\it ASTROSAT}}
\newcommand\xmm{{\it XMM-Newton}}

\newcommand\s{{\rm~s}}

\newcommand\kev{{\rm~keV}}

\newcommand\xiunit{\ifmmode {\rm~erg\s}$^{-1}$ \else ~erg~cm~s$^{-1}$\fi}
\newcommand\kms{\ifmmode {\rm~km\ s}$^{-1}$ \else ~km s$^{-1}$\fi}
\newcommand\Hunit{\ifmmode {\rm~km\ s}$^{-1}$\ {\rm Mpc}$^{-1}$
        \else ~km s$^{-1}$ Mpc$^{-1}$\fi}
\newcommand\ctssec{\ifmmode {\rm~count\ s}$^{-1}$ \else ~count s$^{-1}$\fi}
\newcommand\ergsec{\ifmmode {\rm~erg\ s}$^{-1}$ \else
        ~erg s$^{-1}$\fi}
\newcommand\funit{\ifmmode {\rm~erg\ s}$^{-1}$\;{\rm cm}$^{-2}$ \else
        ~ergs s$^{-1}$ cm$^{-2}$\fi}
\newcommand\phflux{\ifmmode {\rm~photon\ s}$^{-1}$\;{\rm cm}$^{-i2}$
        \else   ~photon s$^{-1}$ cm$^{-2}$\fi}
\newcommand\efluxA{\ifmmode {\rm~erg\ s}$^{-1}$\;{\rm cm}$^{-2}$\;{\rm
        \AA}$^{-1}$ \else ~erg s$^{-1}$ cm$^{-2}$ \AA$^{-1}$\fi}
\newcommand\efluxHz{\ifmmode {\rm~erg\ s}$^{-1}$\;{\rm cm}$^{-2}$\;{\rm
        Hz}$^{-1}$ \else ~erg s$^{-1}$ cm$^{-2}$ Hz$^{-1}$\fi}
\newcommand\cc{\ifmmode {\rm~cm}$^{-3}$ \else cm$^{-3}$\fi}
\newcommand\FWHM{\ifmmode {\rm~FWHM} \else ${\rm~FWHM}$\fi}
\newcommand\Msun{\ifmmode M_{\odot} \else $M_{\odot}$\fi}
\newcommand\Lsun{\ifmmode L_{\odot} \else $L_{\odot}$\fi}

\newcommand\hbeta{\ifmmode {\rm H}\beta \else H$\beta$\fi}
\newcommand\Kalpha{\ifmmode {\rm K}\alpha \else K$\alpha$\fi}
\newcommand\nh{\ifmmode N_{\rm H} \else N$_{\rm H}$\fi}

\makeatletter

\newcommand{\Rmnum}[1]{\expandafter\@slowromancap\romannumeral #1@}
\makeatother

\title [Correlated variability of NGC~4593] {Correlated X-ray/UV/optical emission and short term variability in a Seyfert~1 galaxy NGC~4593}

\author [Main Pal \& S. Naik] { Main Pal\thanks { Email: mainpal@prl.res.in}  \& Sachindra~Naik \thanks{} \\
Astronomy and Astrophysics Division, Physical Research Laboratory, Ahmedabad - 380009, India}

\begin{document}

\pagerange{\pageref{firstpage}--\pageref{lastpage}} \pubyear{}
\date{\today}
\maketitle
\begin{abstract}

We present a detailed multi-frequency analysis of an intense monitoring programme of Seyfert~1 galaxy NGC~4593 over a duration of nearly for a month with \swift{}~observatory. We used 185 pointings to study the variability in six ultraviolet/optical and two soft (0.3-1.5 keV) and hard X-ray (1.5-10 keV) bands. The amplitude of the observed variability is found to decrease from high energy to low energy (X-ray to optical) bands. Count-count plots of ultraviolet/optical bands with hard X-rays clearly suggest the presence of a mixture of two major components: (i) highly variable component such as hard X-ray emission and (ii) slowly varying disc-like component. The variations observed in the ultraviolet/optical emission are strongly correlated with the hard X-ray band. Cross-correlation analysis provides the lags for the longer wavelengths compared to the hard X-rays. Such lags clearly suggest that the changes in the ultraviolet/optical bands follow the variations in the hard X-ray band. This implies the observed variation in longer wavelengths is due to X-ray reprocessing. Though, the measured lag spectrum (lag vs. wavelength) is well described by $\lambda^{4/3}$ as expected from the standard disc model, the observed lags are found to be longer than the predicted values from standard disc model. This implies that the actual size of the disc of NGC~4593 is larger than the estimated size of standard thin disc as reported in AGN such as NGC~5548, Fairall~9.   

\end{abstract}


\section{Introduction}
Active galactic nuclei (AGN) are normally considered to have accreting supermassive black holes (SMBHs) at the heart of the host galaxies. The radiation from these AGN covers almost entire range of electromagnetic spectrum -- starting from X-ray to Radio bands. The X-ray continuum emission from these objects are considered to be dominated by the power law model which is known to be due to the inverse Compton scattering of soft photons in an optically thin i.e., $\tau\sim1$ and hot electron plasma i.e., $kT_{e}\sim100$keV \citep{1980A&A....86..121S,1991ApJ...380L..51H}. The soft photons observed in these sources are believed to be emitted from the accretion disc around the SMBH. The accretion disc, as proposed by \citet{1973A&A....24..337S}, is assumed to be geometrically thin and optically thick. In the accretion disc, the energy loss due to the viscous heating is being emitted as blackbody emission corresponding to the temperature at a given radius. The temperature at a certain distance from the center is inversely proportional to the three--fourth power of the distance. Depending on the temperature, the disc emits radiation in the range of ultraviolet (UV) to optical bands \citep{1999PASP..111....1K,2014ARA&A..52..529Y}. However, sometimes the inner disc becomes sufficiently hot and partly contributes at the soft X-ray band (e.g., \citealt{1996A&A...305...53B, 1999ApJS..125..317L}). 
        
In Seyfert galaxies, radiation emitted from the accretion disc is dominated in UV band. However, due to large Galactic absorption along the line of sight, the UV radiation peak is hardly detectable. The observed UV/optical emission show different variabilities on various timescales from days to years for $10^{6-9}~M_\odot$ black hole mass range  (e.g., \citealt{2008MNRAS.389.1479A, 2011A&A...534A..39M, 2014MNRAS.444.1469M,2014ApJ...788...48S, 2016AN....337..500M}). However, the cause of observed variabilities in UV/optical emission from the accretion disc is not very clear. The observed UV/optical emission variabilities in Seyfert galaxies are thought to be associated with the fluctuation in the mass accretion \citep{2008ApJ...677..880M,2008MNRAS.389.1479A}. However, such interpretation is unable to explain the observed disc variability on short (hours to days) timescale. This variability timescale is very much shorter compared to the expected timescale of the density fluctuations in the accretion flow. The variations observed in X-ray emission are found to lead the variations in UV/optical band \citep{2014ApJ...788...48S,2014MNRAS.444.1469M}. This is difficult to explain by using the hypothesis that the cause of UV/optical variation is associated with accretion flow fluctuation. 

  \citet{1991ApJ...371..541K} suggested that the fluctuations occurring in the UV/optical radiation are delayed compared to the X-ray emission if the variations in the 
X-ray band lead the changes in UV/optical band. This occurs when the X-ray radiation from the compact corona around the SMBH illuminates the optically thick disc and gets absorbed and reprocessed into the UV/optical radiation. The predicted delay in the reprocessed UV/optical radiation with respect to the X-ray radiation varies as the 4/3rd power of wavelength i.e. $\tau\propto\lambda^{4/3}$ \citep{1999MNRAS.302L..24C, 2007MNRAS.380..669C}. Such prediction of lags have been detected in the multi-frequency study of several AGN \citep{2007MNRAS.380..669C,2010MNRAS.403..605B,2014MNRAS.444.1469M,2016MNRAS.456.4040T,2015ApJ...806..129E,2016ApJ...821...56F,2017MNRAS.464.3194B,2017ApJ...840...41E}. On the other hand, \citet{2008RMxAC..32....1G} found that the fluctuations in the UV/optical radiation are independent of the variations in X-rays. It is proposed that the origin of these fluctuations in both the bands are likely to be local to the disc. Occasionally, rapid changes are found in the optical radiation compared to that of the X-ray radiation suggesting distinct regions for their origin (e.g., NGC~3783: \citealt{2009MNRAS.397.2004A}).
 
Multi-frequency observation campaign of AGN provide an opportunity to investigate different regions of the central engine. Till the launch of \xmm{}, only a few contemporaneous long and intensive multi-frequency campaigns were carried out. Monitoring programs with the space-based \rxte{} observatory and ground-based telescopes were carried out for a few AGN (e.g., \citealt{2002AJ....124.1988M, 2008MNRAS.389.1479A, 2009MNRAS.397.2004A,2010MNRAS.403..605B}). At present, there are excellent space-based observatories i.e., \xmm{},~\swift{} and \astrosat~ which have the capabilities to carry out simultaneous multi-band observations in the UV/optical and X-ray bands. A number of Seyfert~1 galaxies have revealed strong correlation between the UV/optical and the X-ray bands \citep{2011A&A...534A..39M,2014MNRAS.444.1469M,2016MNRAS.456.4040T,2015ApJ...806..129E,2016ApJ...828...78N,2017MNRAS.464.3194B,2017arXiv170705536L,2017ApJ...840...41E, 2017MNRAS.466.1777P, 2017ApJ...835...65S, 2016ApJ...828...78N,2016MNRAS.459.3963C,2016AN....337..500M,2016ApJ...821...56F}. In some cases, a relatively moderate correlation have been observed between UV/optical and X-ray bands \citep[e.g., NGC~7469:][]{1998ApJ...505..594N}. However, a few AGN have shown no clear connection in the UV/optical and the X-ray emission \citep{2002AJ....124.1988M, 2016MNRAS.457..875P}. Thus, the correlation and variations between the UV/optical emission and the X-ray emission are complex and their intensive exploration is required. 

To investigate the cause of variation in emitted radiation from different parts of the accretion disc, multi-band monitoring observations of Seyfert~1 galaxy NGC~4593 with \swift{} observatory have been used in the present study. This AGN is a barred Seyfert~1 galaxy, classified as Hubble type SBb, located at a redshift of $z=0.009$ \citep{1992ApJS...83...29S}. This AGN is mildly X-ray luminous with $2-10keV$ luminosity $L_{2-10keV}\sim10^{42}~\rm erg~s^{-1}$ \citep{2013MNRAS.435.3028E} and harbors a supermassive black hole of mass $\sim(1-10)\times10^6~M_{\odot}$ \citep{2006ApJ...653..152D, 2004ApJ...613..682P, 2003ApJ...585..121O,2000ApJ...543L...5G}.  NGC~4593 has been found to be highly variable in X-ray, UV, optical and infrared (IR) bands, suggesting the emitting region to be a compact source at the centre \citep{1995MNRAS.274....1S,2016MNRAS.463..382U}. Previous studies revealed the absence of broad iron K$\alpha$ line in the X-ray spectrum, instead two narrow lines at 6.4 keV and 6.97 keV have been detected \citep{2007ApJ...666..817B, 2009ApJ...705..496M}. \citet{2009ApJ...705..496M} suggested that the accretion disc may be truncated at some inner region with radiatively inefficient flow while the outer disc remains radiatively efficient. However, \citet{2004MNRAS.352..205R} discussed the lack of relativistic iron K$\alpha$ line is possibly due to the low iron abundance, highly ionized surface and high inclination of the disc. They inferred that the radiatively efficient accretion disc is consistent and can be extended down to the inner most stable radius. In this work, we explore various regions of the accretion disc and its coupling with the X-ray continuum by using multi-frequency observations with \swift{}~observatory. 

In the paper, we discuss the observation and the data reduction in Section~2. UV/optical and X-ray light curves analysis and related cross-correlation studies are described in Section~3. We list our findings and discuss the results in Section~4.

\section{Observations and data reduction}

We used \swift{}/Ultraviolet/Optical Telescope (UVOT) \citep{2005SSRv..120...95R} and \swift{}/X-ray Telescope (XRT) \citep{2005SSRv..120..165B} archival data of the AGN from 2016 July 13 to 2016 August 5 in our analysis. Both the instruments are simultaneously used during observations providing coverages in X-ray and UV/optical bands. 

\begin{figure*} 
\includegraphics[scale=0.55]{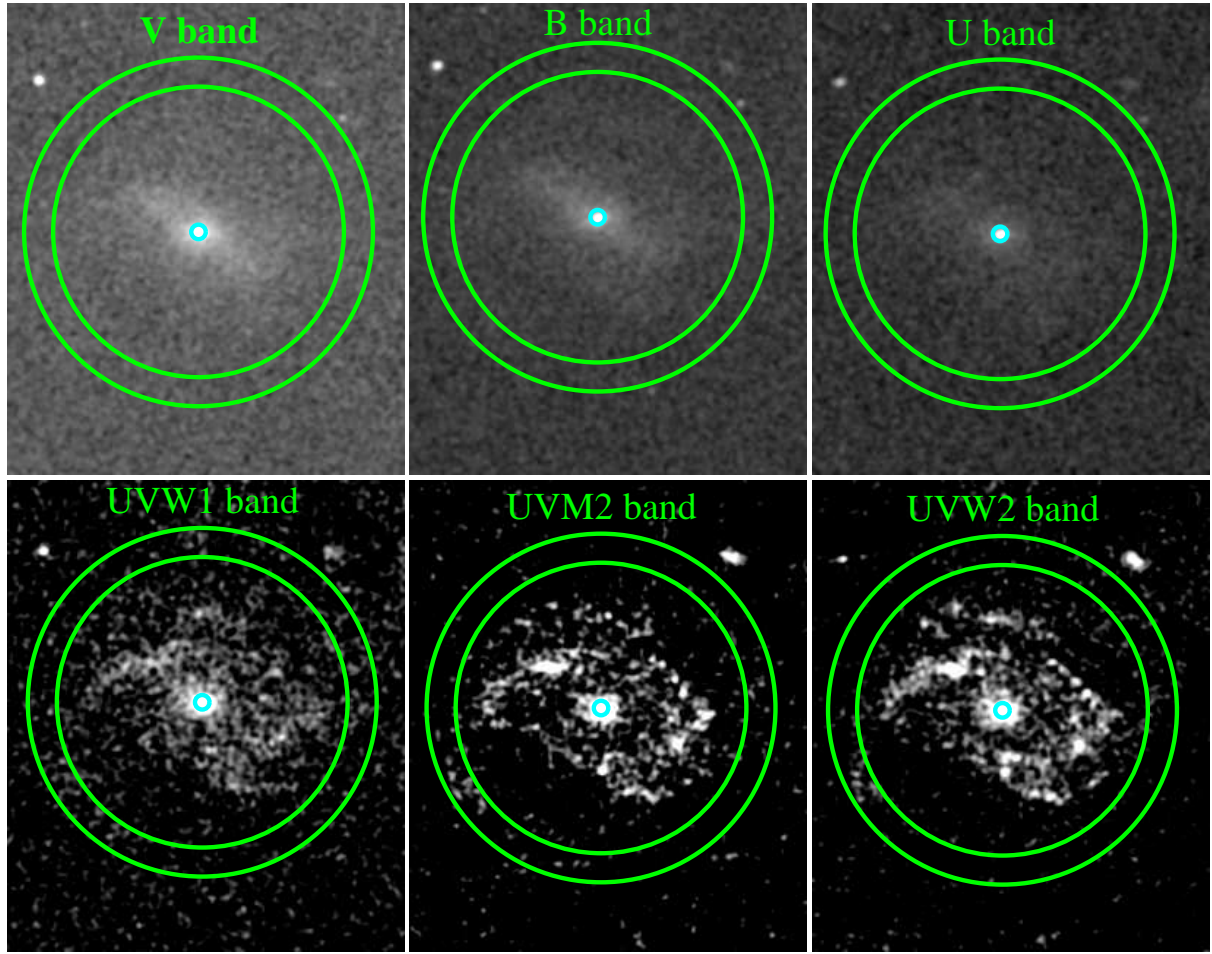}
\caption{ In all the UV/optical bands, the nucleus is located at 
the centre. The extended emission with bar-like structure is clearly visible in 
the optical bands (top panels) while the arm like structure of the galaxy NGC~4593 
appears only in UV bands (bottom panels). Source and background regions used to 
extract UV/optical light curves from an epoch of observation (Image shown for ID :
00092353198) are shown in each panel. Upper panels show the selection of regions 
from V to U-band whereas lower panels show for UVW1 to UVW2 bands. }
\label{obs_net1} 
\end{figure*}

\begin{figure*} 
\includegraphics[scale=0.45]{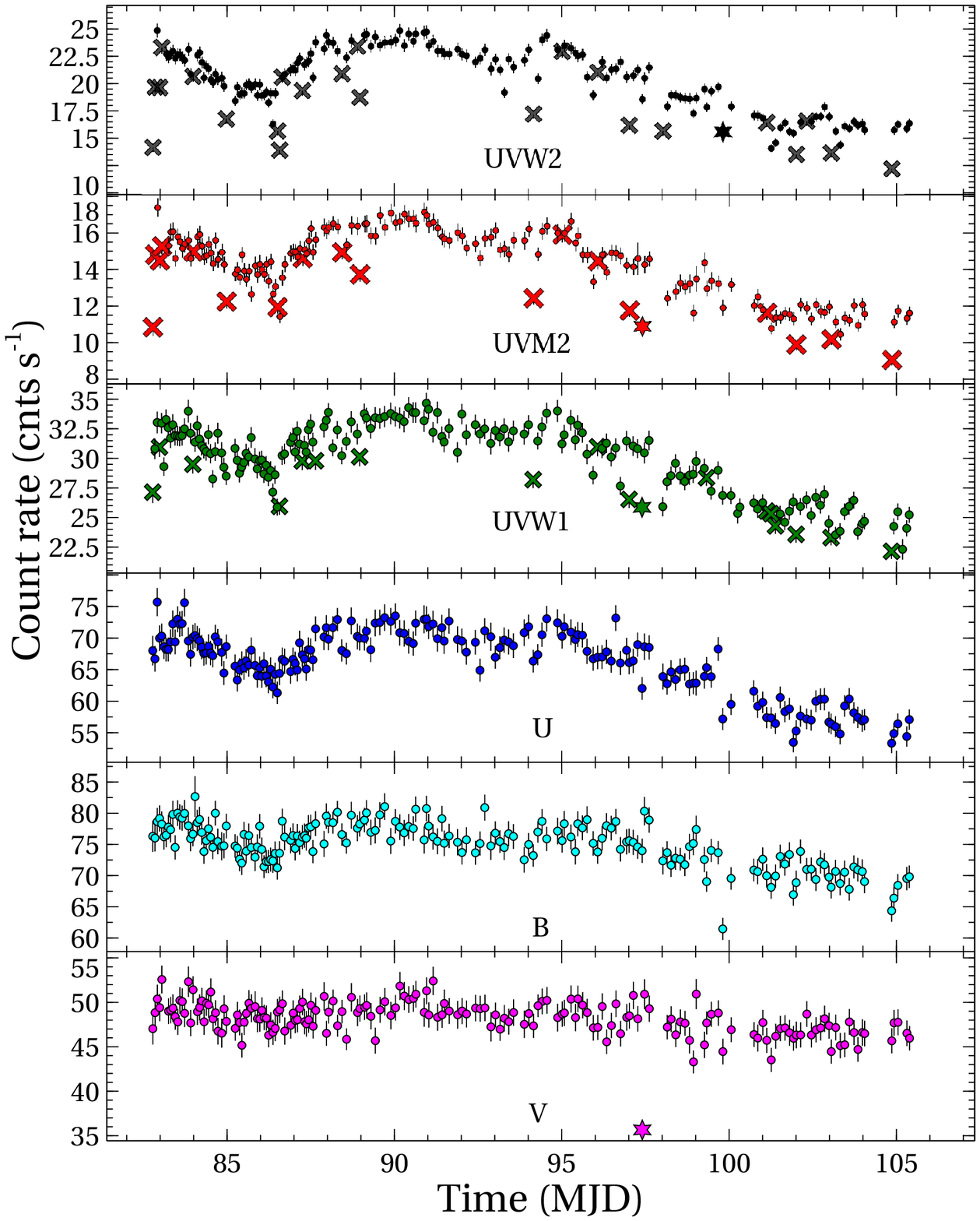}
\includegraphics[scale=0.5]{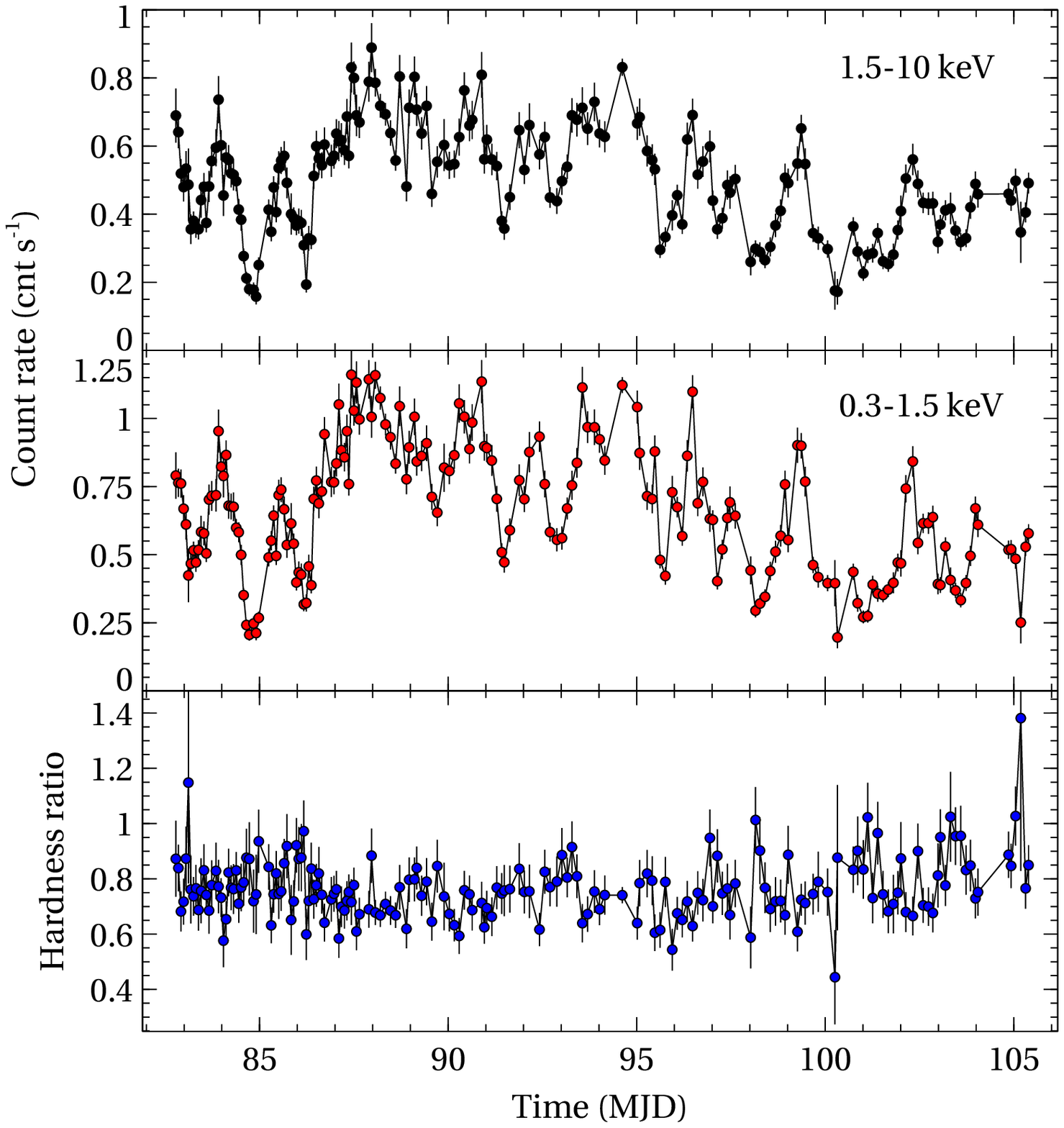}
\caption{Left panels : UV (UVW2, UVM2, UVW1 and U bands) and optical (B and V bands) light curves, simultaneous to X-ray light curves are presented from top to bottom panels. Right panels : Light curves in $1.5-10$\kev~(top panel), $0.3-1.5$\kev~(middle panel) ranges and hardness ratio i.e., $1.5-10$\kev/$0.3-1.5$\kev(bottom panel) are shown. The time axis is in MJD-57500.}
\label{uvobs1} 
\end{figure*}

Data obtained from UVOT observations, contemporaneous to that of XRT, were reduced by following standard procedure. We used sky image files, free from any change in the source position, to get the count rate for each filter of every observation. We summed available multiple frame exposures by using {\tt UVOTIMSUM} to increase the signal to noise ratio. A circular region of radius 5 arcsec centred at source coordinate was selected to extract the spectrum for all filters of each observation. Annular regions with inner and outer radii of 100 and 120 arcsec, respectively, with source at the centre were selected for background estimation for all the observations (see  Fig.~\ref{obs_net1}). The background
region is carefully selected to avoid the contribution from extended components in the 
galaxy. {\tt UVOTSOURCE} package was used internally by {\tt UVOT2PHA} task to compute the source count rate by using latest calibration files\footnote{https://heasarc.gsfc.nasa.gov/ftools/caldb/help/uvot2pha.html}. Following above procedure, source and background spectral products were extracted for each observation. Background corrected source count rates for all the filters were estimated for every observation. We also checked small scale sensitivity (SSS) by running {\tt UVOTSOURCE} on sky images of all the filters using latest SSS file\footnote{https://swift.gsfc.nasa.gov/analysis/uvot\_digest/sss\_check.html}. The data 
points which landed on the low sensitivity patches of the CCD are marked with crosses 
and are shown in the left panels of Fig.~\ref{uvobs1}. Apart from this, there are very 
few data points, marked with asterisk in the left panels of Fig.~\ref{uvobs1}, which 
show $\ge$15\% variability compared to the local mean. This variability might be caused
due to the tracking of telescope as quoted in \citet{2014MNRAS.444.1469M}. We excluded 
all these data points marked with crosses and asterisks in the figure from our analysis.
This resulted in a total of 160 to 184 usable data points (average count rates) for UVOT filters.

As described in \citet{2007A&A...469..379E,2009MNRAS.397.1177E}, “{\it Swift}/XRT 
product generator\footnote{http://www.swift.ac.uk/user\_objects/}” online software 
package was used to extract source light curves. On-target exposure of every pointing 
was about 1 ks (Table~\ref{obs_log}). For each monitoring epoch, average count rate 
over entire duration of observation was extracted from photon counting (PC) mode of 
XRT. We found a total of 185 usable data points for our study (see right panel of 
Fig.~\ref{uvobs1}). For spectral study, we used {\tt XRTPIPELINE} 
to extract event files incorporating the latest calibration data base. This tool 
selects a circular region of radius 20 pixels ($\sim$47 arcsec) centred at the 
source co-ordinates to extract source spectrum. We also created background spectrum 
using an annular region with inner and outer radii of 20 and 45 pixels, respectively, 
centred at the source co-ordinates. Since PC mode data sets are affected by pile-up 
in our case, we removed the circular core of point spread function (PSF) of radius 
in the range 2 to 4 pixels as recommended by instrument team to mitigate the effect. 
We also generated the effective area file by using {\tt XRTMKARF}. We grouped the 
source spectrum to get a minimum of 15 counts per bin to make use of $\chi^{2}$ 
minimization technique.

\begin{table}
  \noindent \begin{centering}
    \caption{Log of observation of NGC~4593 with \swift{}~XRT/UVOT} \label{obs_log}
    \begin{tabular}{lr}
      \hline 
Observation ID   & 00092353001--00092353201\\
Date of Observations  & 2016 July 13 - 2016 August 5\\
MJD                   & 57582.8 - 57605.4\\
No. of IDs for XRT    &185\\
No. of IDs for UVOT   &160-184\\
       \hline
    \end{tabular}
    \par\end{centering}
\end{table}

\section{Data analysis}

\subsection{Count--Count Correlation with Positive Offset (C3PO)}
\begin{figure} \centering
\includegraphics[scale=0.5]{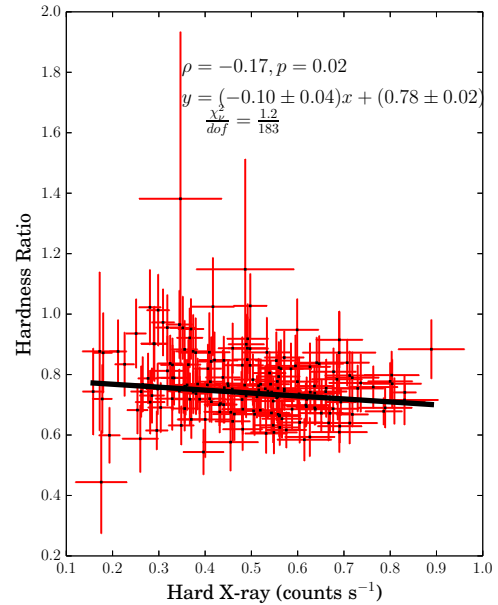}
\caption{The correlation between hardness ratio and hard X-ray count rate is shown. This provides an anti-correlation which suggests that this source is softer when brighter.}
\label{sw_obs1} 
\end{figure}

The $1.5-10$\kev~and $0.3-1.5$\kev~bands light curves for the entire duration of monitoring campaign were created by using observed average count rates for each epoch of observation. In our analysis, the $1.5-10$\kev~band stands for the hard X-ray band where power-law continuum is thought to be the primary component. The $0.3-1.5$\kev~band stands for the soft X-ray band which may contain a complex mixture of multiple spectral components such as due to warm absorption along the line of sight, blurred reflection from partially ionized disc and possible intrinsic disc Comptonization. In right panel of Fig.~\ref{uvobs1}, the hard X-ray light curve, the soft X-ray~light curve and their ratio (1.5-10 keV/0.3-1.5 keV) are shown for all observations from 2016 July 13 to 2016 August 5. It can be seen from the figure that the both soft and hard X-ray emission vary about five times on a few week timescale and vary by a factor of $\sim2$ on short (a few hours) timescale. The hardness ratio appears almost constant during the entire monitoring duration while it seems to vary rapidly on a few hours timescale. Similarly, we found that UV and optical emission are also highly variable and each light curve appears to be correlated to each other. Further, the hardness ratio and hard X-ray count rate
seem anti-correlated suggesting this AGN retains its nature being softer when brighter as 
shown in Fig.~\ref{sw_obs1} \citep{2016MNRAS.463..382U}.

\begin{figure*} \centering
\includegraphics[scale=0.9]{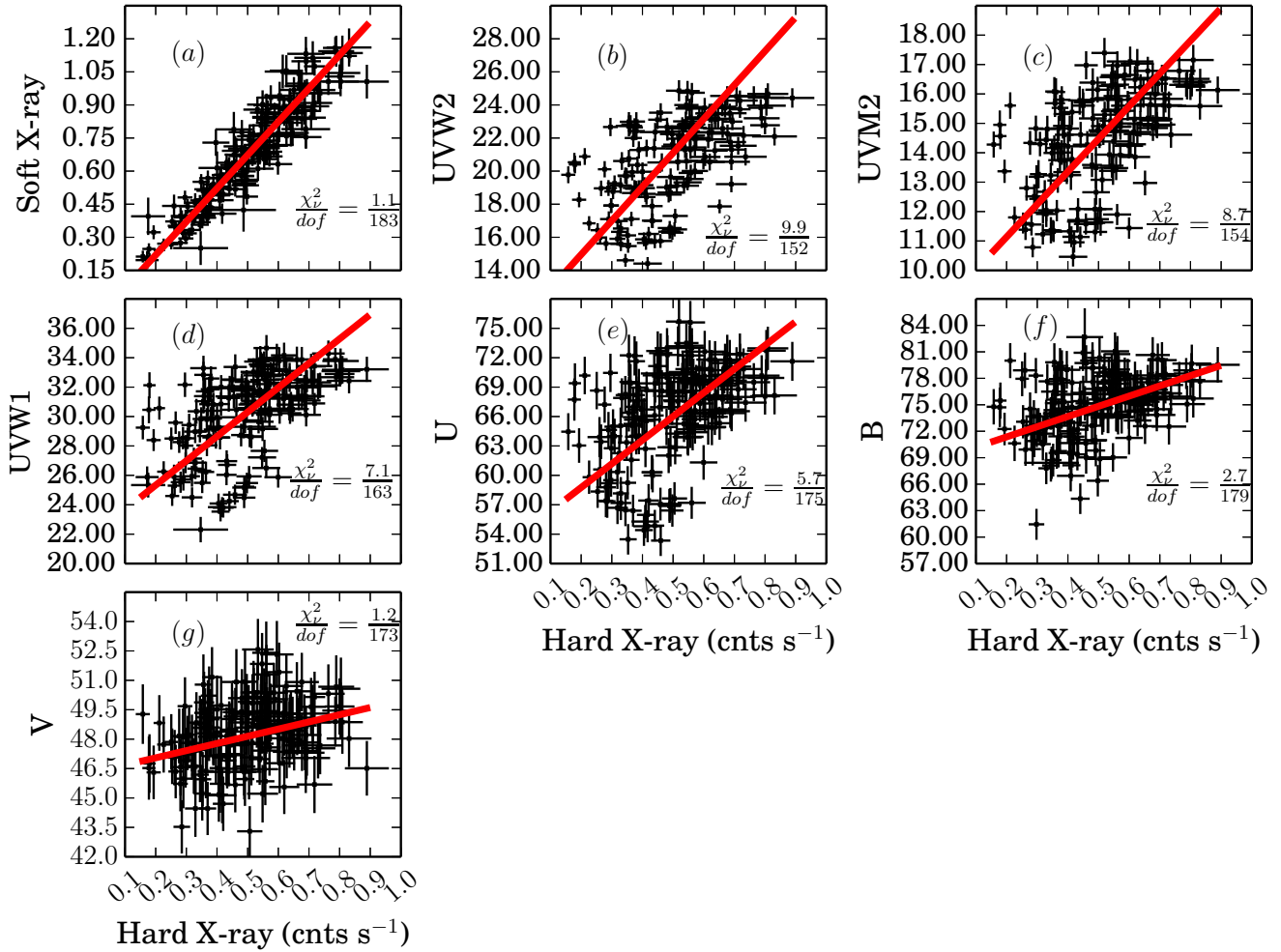}
\caption{Count-count plots between different bands are shown. The linear fittings (y=m$\times$x+c, where $m$ and $c$ are slope and offset, respectively) to the correlation between soft bands (soft X-ray to V bands) and the hard X-ray band are shown as solid lines. The best-fit linear equation obtained from modeling are (a) $y=(1.51\pm0.03)x+(-0.08\pm0.02)$ (b) $y=(20.5\pm1.8)x+(10.9\pm0.8)$ (c) $y=(11.1\pm1.0)x+(8.9\pm0.5)$ (d) $y=(16.6\pm1.5)x+(22.0\pm0.7)$ (e) $y=(24.2\pm0.2.6)x+(53.9\pm1.3)$ (f) $y=(11.6\pm1.6)x+(69.0\pm0.8)$ (g) $y=(3.7\pm0.8)x+(46.3\pm0.4)$. In each panel, the count rates used are in the unit of counts $s^{-1}$.}
\label{corr} 
\end{figure*}

We fitted each pair of soft band (soft X-ray, UVW2, UVM2, UVW1, U, B and V) and hard X-ray band with a line $y=m\times~x+c$, where $m$ and $c$ are slope and offset. The slope $m$ tells about how strongly $y$ depends on $x$, while the offset $c$ provides information about the slowly varying component as discussed by \citet{ 2013PASJ...65....4N, 2013ApJ...771..100N, 2011PASJ...63S.925N}. This method is known as count-count correlation with positive offset (C3PO). Such a method was first developed by \citet{2001MNRAS.321..759C} in a binary system Cygnus~X-1 to decompose the stable and variable components. Later, \citet{2003MNRAS.342L..31T} applied above method on AGN. We used hard X-ray band as the abscissa (e.g., $x$) and any soft band as the ordinate (e.g., $y$). After linear function fitting to the pair of a soft band and hard X-ray band, we found that the soft X-ray consists of a negative offset ($-0.08\pm0.02$) while other UV/optical bands have positive offset (e.g., $\sim9-57$). The marginal negative offset for soft X-ray band suggests that this band is possibly affected by absorption as reported in previous studies (e.g., \citealt{2016MNRAS.463..382U, 2013MNRAS.435.3028E}). The positive offset for UV/optical emission hints a presence of weakly varying component such as disc emission. The positive slopes (e.g, $\sim1.5-24$) indicate a strong relationship between soft bands and hard X-ray band as shown in count-count plots of Fig.~\ref{corr}). However, the linear fit is not enough to describe all bands due to the scatters around the fitted line except the soft X-ray and V bands (see Fig.~\ref{corr}). 

\begin{figure*}  
\includegraphics[scale=0.9]{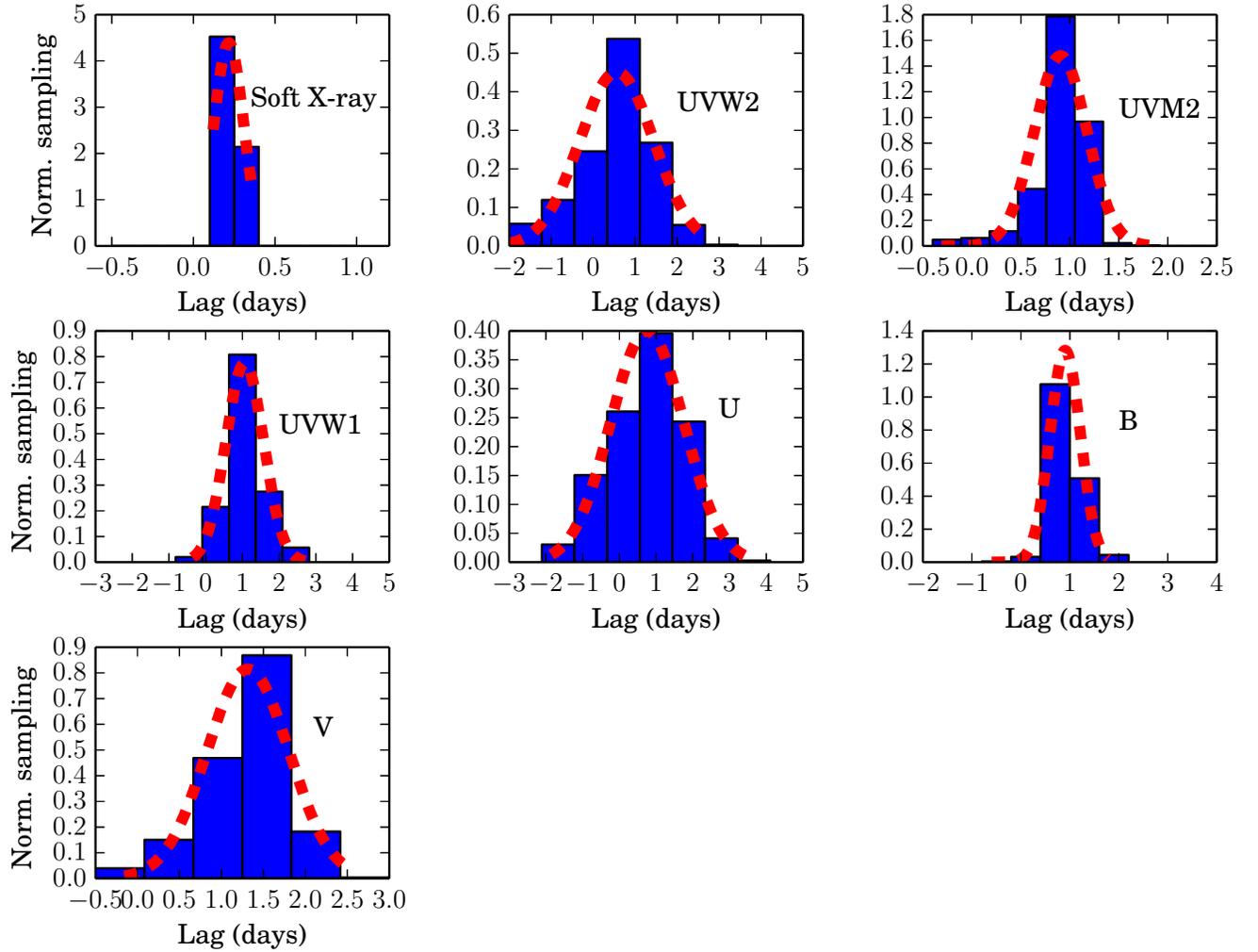}
\caption{Probability distribution of lags for soft bands with respect to 1.5-10 keV band by using {\tt JAVELIN} code. The dashed-red curve in each panel shows the Gaussian representation of probability distribution of the observed lag.}
\label{javelin}
\end{figure*}

The possible cause of these scatters can be due to the contributions from the broad 
line regions (BLR) and narrow line regions (NLR) of the host galaxy. An estimate of 
the fractional contribution due to the emission lines from the BLR and NLR can be 
made by taking the ratio between the net equivalent widths of the lines in the band 
and the full width at half maximum (FWHM) of the filter. We used equivalent widths of 
all narrow and broad emission lines from \citet{2001AJ....122..549V} in UVW2, UVM2, 
UVW1, U, B and V bands and obtained the net equivalent widths in each band. We also 
obtained the FWHM for each band from \citet{2008MNRAS.383..627P}. We then divided 
the net equivalent width of all the lines in each band by the respective FWHM. We 
found the resultant fraction to be $\sim$3.5\% in V band, $\sim$9\% in B band, 
$\sim$1.2\% in U band, $\sim$5.5\% in UVW1 band, $\sim$1.6\% in UVM2 band and 
$\sim$4.1\% in UVW2 band. These fractions clearly suggest that the total emission 
in various bands are affected from BLR/NLR components of the host galaxy.

\subsection{Time lag estimation}
We used Pearson's correlation coefficient~`$\rho$' to quantify the strength of inter-band correlation (soft band and hard X-ray light curves). We also determined the significance of the strength of correlation between longer wavelength bands and hard X-ray band. We obtained the correlation coefficient `$\rho$' to be within $\sim\rho=0.29$ and $\rho\sim0.94$ from optical/UV to soft X-ray bands with very low probability $p$ ($\sim p=1.6\times10^{-4}-2.8\times10^{-84}$ for V to soft X-ray band) which occurs by chance. This can be seen clearly in Fig.~\ref{corr} from top to bottom for soft X-ray to V band. Details about the method of estimating correlation coefficient and probability are described in \citet{2017MNRAS.466.1777P}.

\begin{figure*} \centering
\includegraphics[height=12.5cm,width=7.3cm]{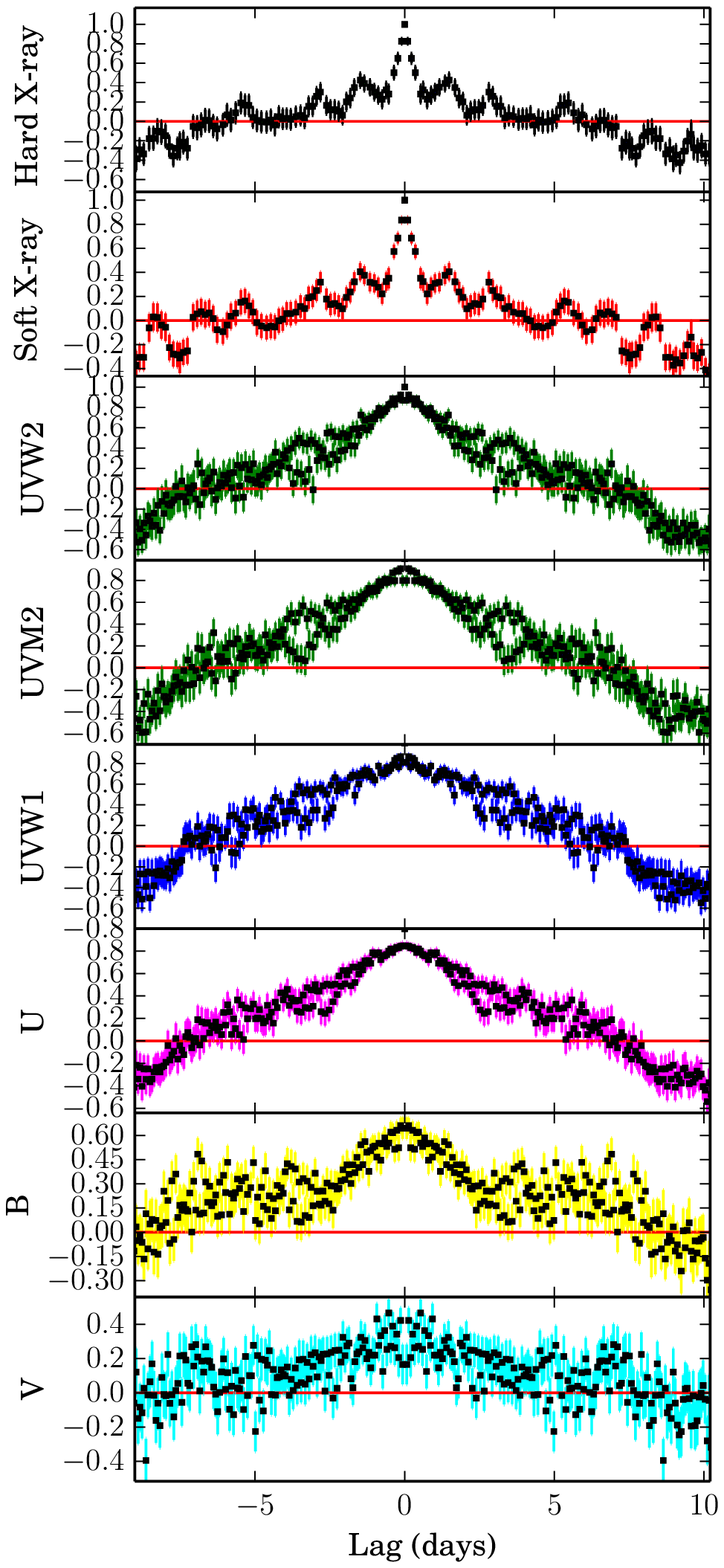}
\includegraphics[scale=0.56]{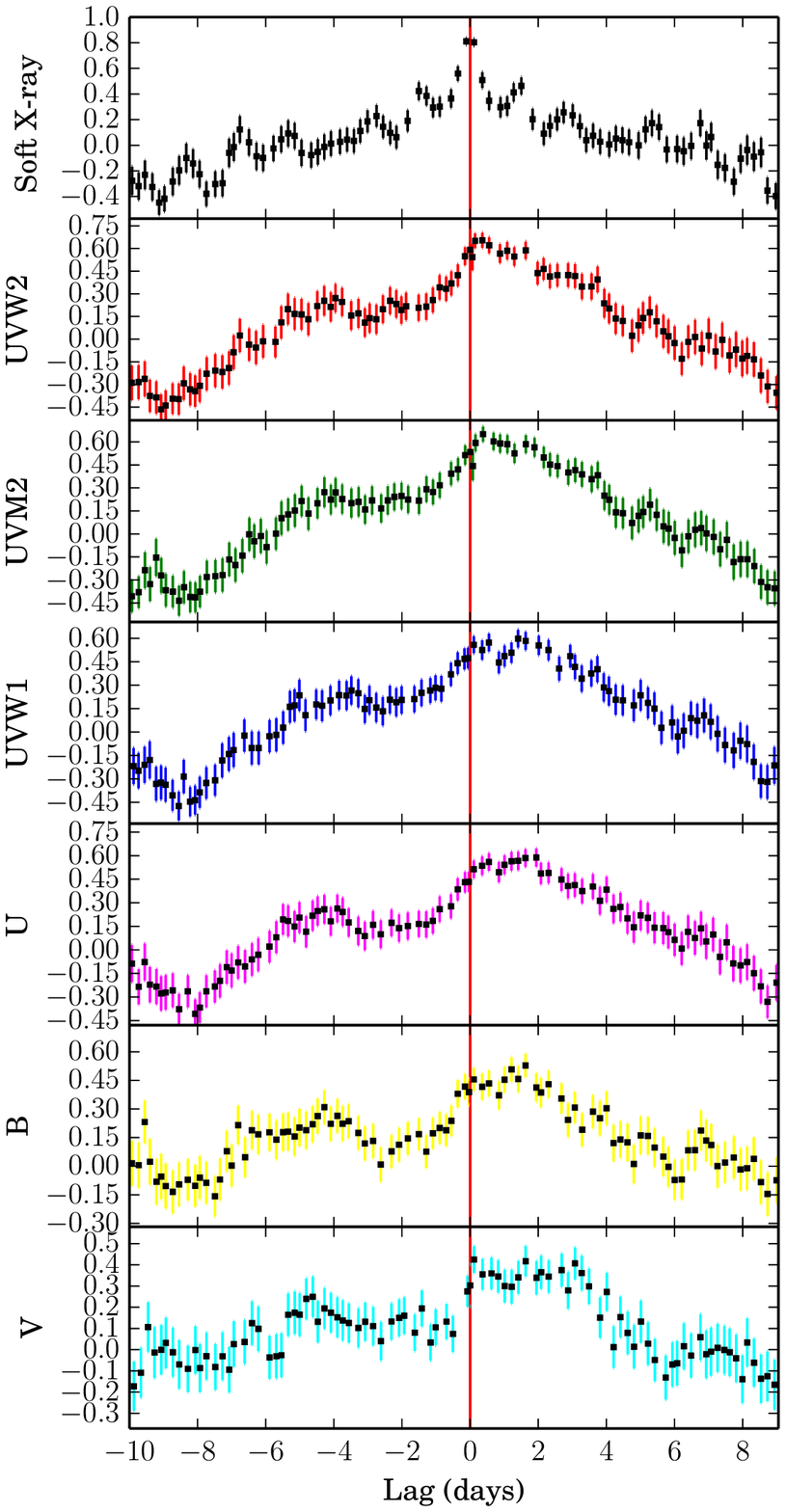}
\caption{Left panel: From top to bottom- auto correlation function (ACF) of hard X-ray to UV/optical light curves using {\tt ZDCF} function. Right panel: From top to bottom- the cross-correlation function (CCF) for the soft X-ray and each UV/optical bands compared to $1.5-10$\kev band. The vertical solid line in the right panels represents zero delay.}
\label{zdcf}
\end{figure*}

We computed the possible lag/lead between soft band and hard X-ray light curves. There are publicly available codes such as {\tt JAVELIN} \citep{2011ApJ...735...80Z, 2013ApJ...765..106Z} and Z-transformed discrete cross-correlation function ({\tt ZDCF}) \citep{1997ASSL..218..163A} which are capable of finding the presence/absence of any timing information between the light curves. {\tt JAVELIN} software is based on damped random walk process which has been commonly found in quasar variability. This phenomenon was first introduced by \citet{2009ApJ...698..895K}. Since then, it has been used to understand the observed variability in the UV/optical bands \citep{2010ApJ...708..927K, 2010ApJ...721.1014M, 2011ApJ...735...80Z, 2013ApJ...765..106Z}. We applied this software by assuming hard X-ray band as the first light curve and the soft bands as the second light curve to determine the lag distribution. The lag distribution between soft bands and hard X-ray band are shown in Fig.~\ref{javelin}. Assuming Gaussian distribution, we estimated lags for soft bands with respect to hard X-ray band and found to be for soft X-ray band -- $0.22\pm0.09$ days, UVW2 band -- $0.56\pm0.88$ days, UVM2 band -- $0.91\pm0.27$ days, UVW1 band -- $1.1\pm0.5$ days, U band -- $0.76\pm1.0$ days, B band -- $0.9\pm0.3$ days and V band -- $1.3\pm0.5$ days. 

We also applied {\tt ZDCF} to assess the presence of time delay between the soft bands and the hard X-ray band. We run the package with minimum eleven data points, which have non-zero lag, along with non-uniform binning between a pair of bands to determine the cross-correlation function (CCF). We also derived the auto-correlation function (ACF) for each band using minimum eleven data points with non-uniform binning. The resulted CCF and ACF are displayed in the right and left panels of Fig.~\ref{zdcf}, respectively. The ACF of each UV/optical band appears broader compared to the X-ray bands. The CCF of soft bands also show broader distribution towards the positive lag. This implies that the longer wavelength is delayed with respect to the hard X-ray band. We, therefore, estimated the lag of soft bands compared to the hard X-ray band by using {\tt PLIKE} software provided by \citet{1997ASSL..218..163A}. To quantify lags between soft bands and hard X-ray band, we considered CCF for all the soft bands in the range of $-2$ to 6 days. Following this procedure, the lags are determined to be -- for soft X-ray band -- $-0.11_{-0.14}^{+0.22}$ days, UVW2 band -- $0.36_{-0.19}^{+0.48}$ days, UVM2 band -- $0.38_{-0.13}^{+0.60}$ days, UVW1 band -- $1.41_{-0.95}^{+0.48}$ days, U band -- $1.94_{-0.75}^{+0.11}$ days, B band -- $1.62_{-0.51}^{+0.21}$ days and V band -- $1.14_{-0.84}^{+2.45}$ days with respect to the hard X-ray band. The estimated lags by using {\tt ZDCF} are found to be comparable to that of by using {\tt JAVELIN} software.

\subsection{Modeling of time lag spectrum}

\begin{figure*}
\centering
\includegraphics[scale=0.42]{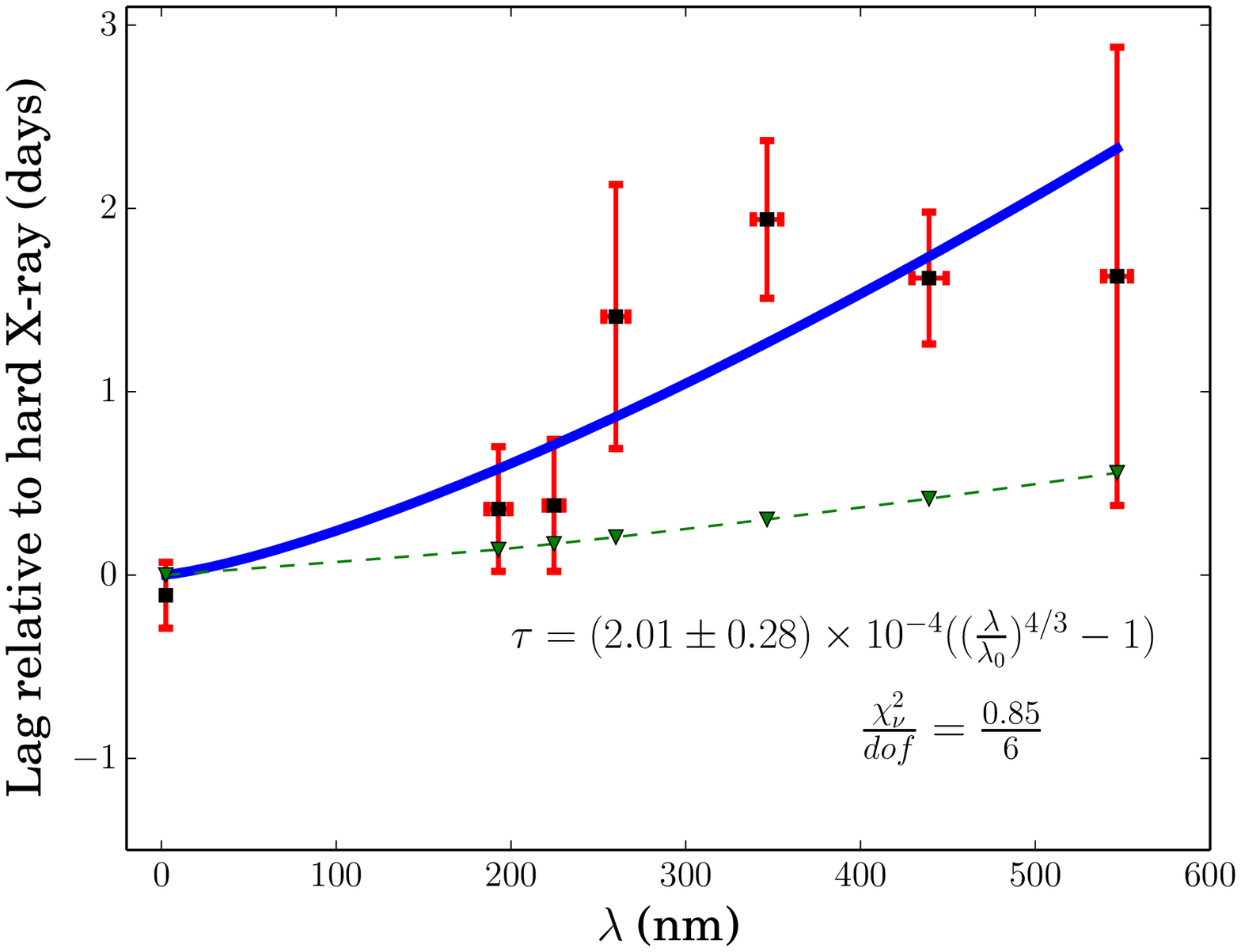}
\includegraphics[scale=0.42]{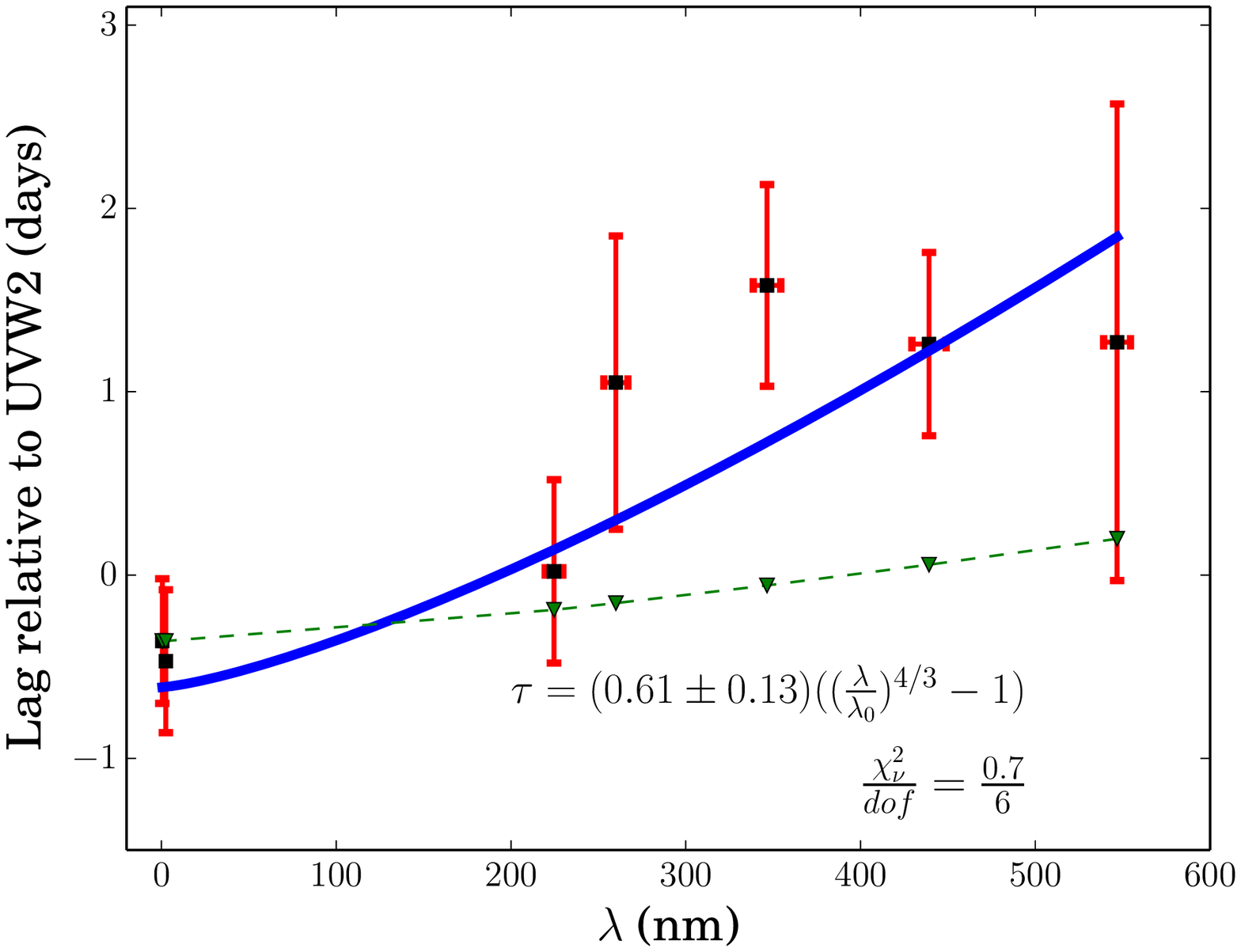}
\caption{ Left panel: The lag spectrum with respect to the hard X-ray band. The fit for the lag spectrum (a) Blue curve- the best-fit power law model ($2.0\pm0.3\times10^{-4}[(\lambda/0.475)^{4/3}-1]$). (b) Green dashed curve : standard accretion thin disc theoretical model. Right panel: The lag spectrum with respect to the UVW2 band. The fit for the lag spectrum (a) Blue curve- the best-fit power law model ($0.6\pm0.1\times10^{-4}[(\lambda/0.475)^{4/3}-1]$). (b) Green dashed curve : standard accretion thin disc theoretical model for NGC~4593. Clearly, the larger lag of U band suggests an extra component of emission possibly emission from broad line region.}
\label{lag_spec}
\end{figure*}

 The central wavelength for each UV/optical band is considered as the mid-point
between the wavelengths at half-maximum and is taken from \citet{2008MNRAS.383..627P}. The 
soft X-ray and hard X-ray bands are expressed in terms of corresponding central wavelengths 
as $2.5\pm1.7~\rm~nm$ and $0.48\pm0.35~\rm nm$, respectively. The derived lags by using 
{\tt ZDCF} and {\tt JAVELIN} softwares for the soft X-ray, UVW2, UVM2, UVW1, U, B and V 
bands are comparable. The lags, determined by {\tt ZDCF}, were used to create the lag 
spectrum where lag is a function of wavelength.  The derived lag spectrum is shown in 
Fig.~\ref{lag_spec}. 

\subsubsection{ Power law model}
 Power law model, normally used for thin disc, is expressed as  

\begin{align}
\label{equ:lag_wv}
\tau = \alpha\left[\left(\frac{\lambda}{\lambda_0} \right)^{\beta}
  -1\right]
\end{align}

where $\tau$, $\alpha$, $\beta$ are time delay for a central wavelength $\lambda$, 
power-law normalization and power-law index, respectively. Here, $\lambda_{0}$ is 
considered as a reference wavelength. According to the standard disc model, the time 
lag varies with 4/3rd power of wavelength. Here, we used the central wavelength of 
hard X-ray band ($\lambda_{0}=0.475~\rm~nm$) as the reference wavelength. We modeled 
the measured lag spectrum with power law model (Eq. ~\ref{equ:lag_wv}) by 
allowing normalization and index to vary (see Eq.~\ref{equ:lag_wv}). In our 
fitting, as the power-law index was not well constrained, we fixed the index at 
4/3. The best-fit model was found to be $\tau = (2.01\pm0.28)\times10^{-4} 
\left[\left(\frac{\lambda}{0.475} \right)^{4/3}-1\right]$ with statistics 
$\chi^{2}_{\nu}/dof=0.85/6$.

\subsubsection{Shakura-Sunyaev accretion disc model}

In the lamppost geometry, the X-ray source is assumed to be at a height $h$ above the disc in the vicinity of the SMBH. Radiation from the X-ray source gets absorbed in the disc which then  re-emitted in longer wavelengths i.e., UV and optical bands. Thus, the thermal emission and the X-ray reprocessed emission from the disc increase the temperature of the disc at a given radius $r$. The following expression includes both these components to give a net temperature $T(r)$ at a radius $r$ \citep{2007MNRAS.380..669C,2000ApJ...535..712B}. 

\begin{align}
\label{Tprofile1}
T(r) = \left(\frac{3GM\dot M}{8\pi \sigma r^3} (1-\frac{r_{in}}{r})^{1/2}+ \frac{(1-A)L_{\rm x}}{4\pi
    \sigma r_{x}^2} ~\rm cos~(\theta_{x})\right)^{1/4}
\end{align}
where $M$, $G$, $\dot M$, $r_{in}$, $r_{x}$, $L_{x}$ and $A$ are the black hole mass, gravitational constant, the mass accretion rate, radius of the innermost stable orbit and the distance of the disc element (i.e., annulus ring) from the X-ray source, luminosity of the source and albedo for X-ray heating, respectively. 
The term $\theta_{x}$ stands for the angle between the normal to the annulus ring of the disc and the line connecting to the annulus ring and the X-ray source. Cosine of $\theta_{x}$ and $r_{x}$ can be expressed as $\frac{h}{r_{x}}$ and  $(h^2+r^2)^{1/2}$, respectively. Assuming height $h$ and inner radius $r_{in}$ to be very small compared to radius $r$, $\frac{cos (\theta_{x})}{r_{x}^{2}}$ and $(1-\frac{r_{in}}{r})^{1/2}$ in Eq.~\ref{Tprofile1} approximately become $\sim\frac{h}{r^{3}}$ and $\sim1$, respectively. $r$ can be represented in terms of $T(r)$, $M$, $\dot{M}$, $h$, $A$ and $L_{x}$. Using Wein's law, temperature $T(r)$ takes the form into the wavelength, $r$ can be expressed as a function of wavelength. Then, $r$ can be written as a product of time delay $\tau$ and light speed $c$. Thus, for a given reference wavelength $\lambda_0$, delay $\tau$ can be expressed as
       
\begin{align}
\label{equ:MmdotLx}
\tau - \tau_0 =
\left(\frac{1}{c}\right)\left(\frac{\lambda_0}{k}\right)^{4/3}\left(\frac{3GM\dot M}{8\pi \sigma } + \frac{(1-A)L_{\rm x} h }{4\pi \sigma}  \right)^{1/3}\nonumber \\
\left[\left( \frac{\lambda}{\lambda_0} \right)^{4/3} - 1  \right].
\end{align}
 where $k$ is $2.9\times10^6$ nm K.

  Eq.~\ref{equ:MmdotLx} allows to estimate the lags based on the standard disc theory. We used $M=1\times10^7~M_{\odot}$ \citep{2006ApJ...653..152D}, $h=6r_{g}$ and $r_{in}=6r_{g}$ and $A=0.2$. The mass accretion rate in the unit of Eddington mass accretion rate $\dot{m_{E}}(=\frac{\dot{M}}{\dot{M_{E}}})=0.04$ and luminosity $L_{x}=10^{43.7}$\ergsec~were also used required to estimate the lags \citep{2009MNRAS.392.1124V}. The lag  $\tau_{0}$ corresponding to the reference wavelength $\lambda_{0}$ would be zero day in the lag calculation from the standard accretion disc model. The theoretically predicted lags are shown as green dashed line in Fig.~\ref{lag_spec}. The filled triangles on the predicted lag curve correspond to the effective wavelength of the respective band. Thus,  the observed lags, marked as filled squares in the figure, are found to be longer than the estimated lags from the Shakura-Sunyaev disc model. 

Assuming face-on accretion disc, the normalization can provide a rough estimate to the size of emitting region of the reference wavelength (e.g., \citealt{2015ApJ...806..129E}). In our case, we assumed the hard X-ray as the reference wavelength. We, therefore, can estimate the size of the X-ray emitting region. Using the best-fit value of $\alpha=2.01\pm0.28\times10^{-4}$, the estimated size of the emitting region is $0.0002\times86400\times3\times10^5 ~\sim10^{7}$ km. For a given mass of black hole $M$, the Schwarzchild radius is $R_{S}\sim2.96(\frac{M}{M_{\odot}}$) km. Using the mass of NGC~4593 ($\sim10^7 \odot{M}$), the size of the regions is found to be $\sim10^{7}$ km as expected. Similarly, the size of UVW2 region is found to be $\sim$530R$_{S}$.

We summarize as follows: the power law model with 4/3rd power of wavelength, as expected from the standard disc, fits the observed lag spectrum. Using the available mass, accretion rate, bolometric luminosity and albedo, theoretically estimated lags are found to be smaller compared to the observed lags. For example, the predicted lag of B band is $\sim0.5$ days whereas the observed lag of B band is $\sim 1.6$ days. This implies that the real accretion disc appears larger in size compared to the predicted size by the standard disc model.

\begin{figure}
\centering
\includegraphics[scale=0.32, angle=-90]{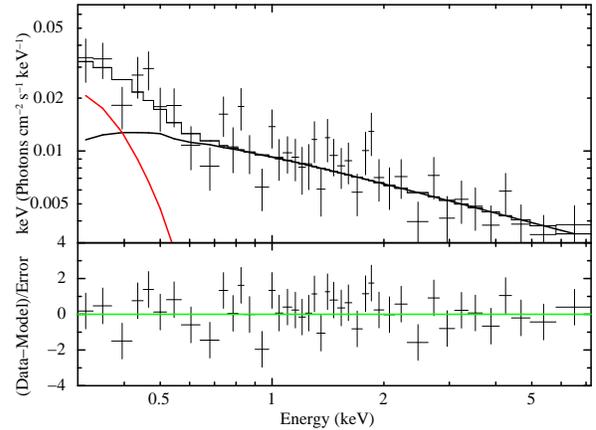}
\caption{00092353119: The best--fit model, data and residuals in $\sigma$ are displayed here. The red solid curve represents the blackbody component used for describing the soft X-ray excess and black solid line exhibits the power--law continuum model.}
\label{bestfit}
\end{figure}

\begin{figure*} 
\includegraphics[scale=0.52]{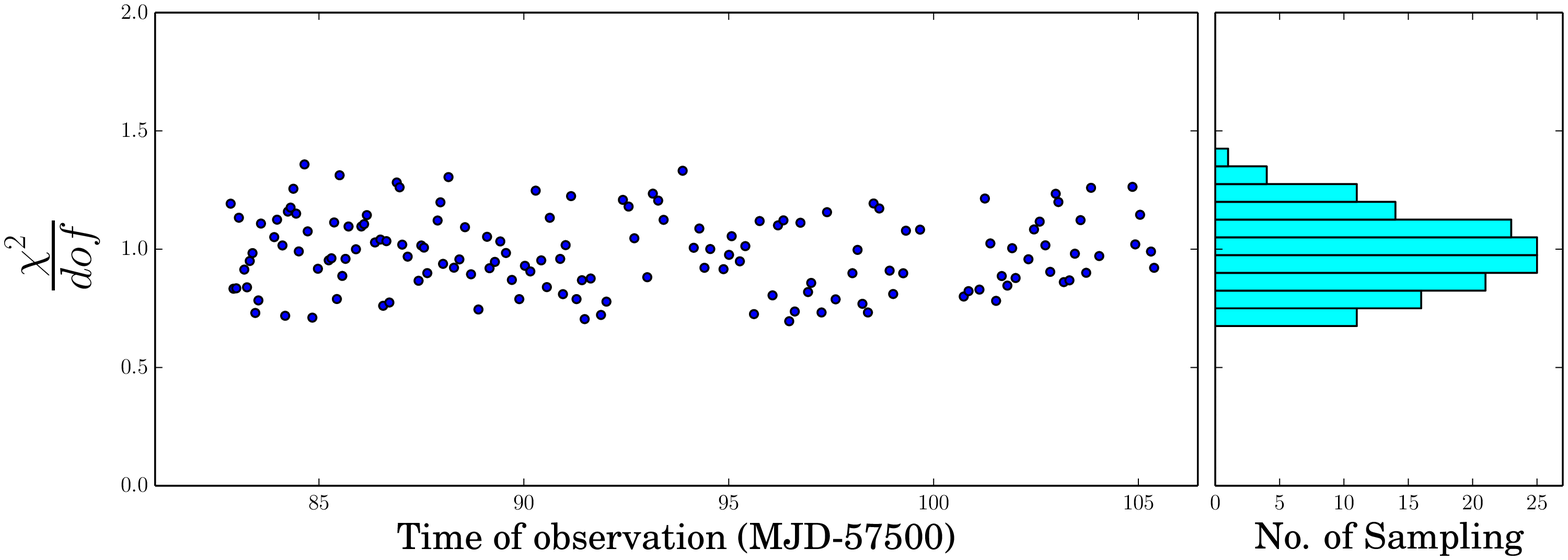}
\includegraphics[scale=0.71]{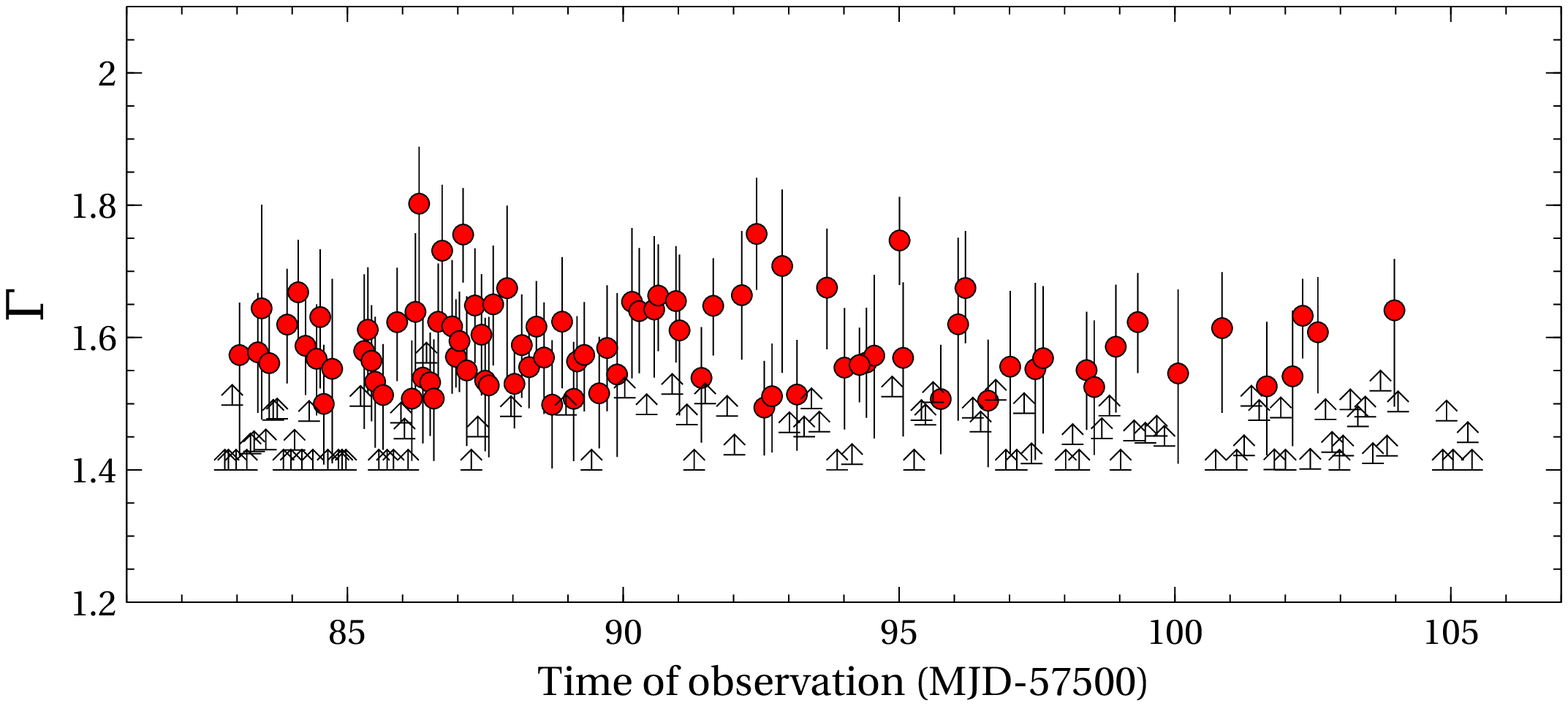}
\caption{The reduced $\chi^{2}$ squared in upper panel for each observation with its MJD. This clearly shows that simple powerlaw+blackbody model is an acceptable model to describe the X-ray band. The time variation of photon index $\Gamma$ is shown in lower panel.}
\label{dist} 
\end{figure*}

\begin{figure*}
\includegraphics[scale=0.42]{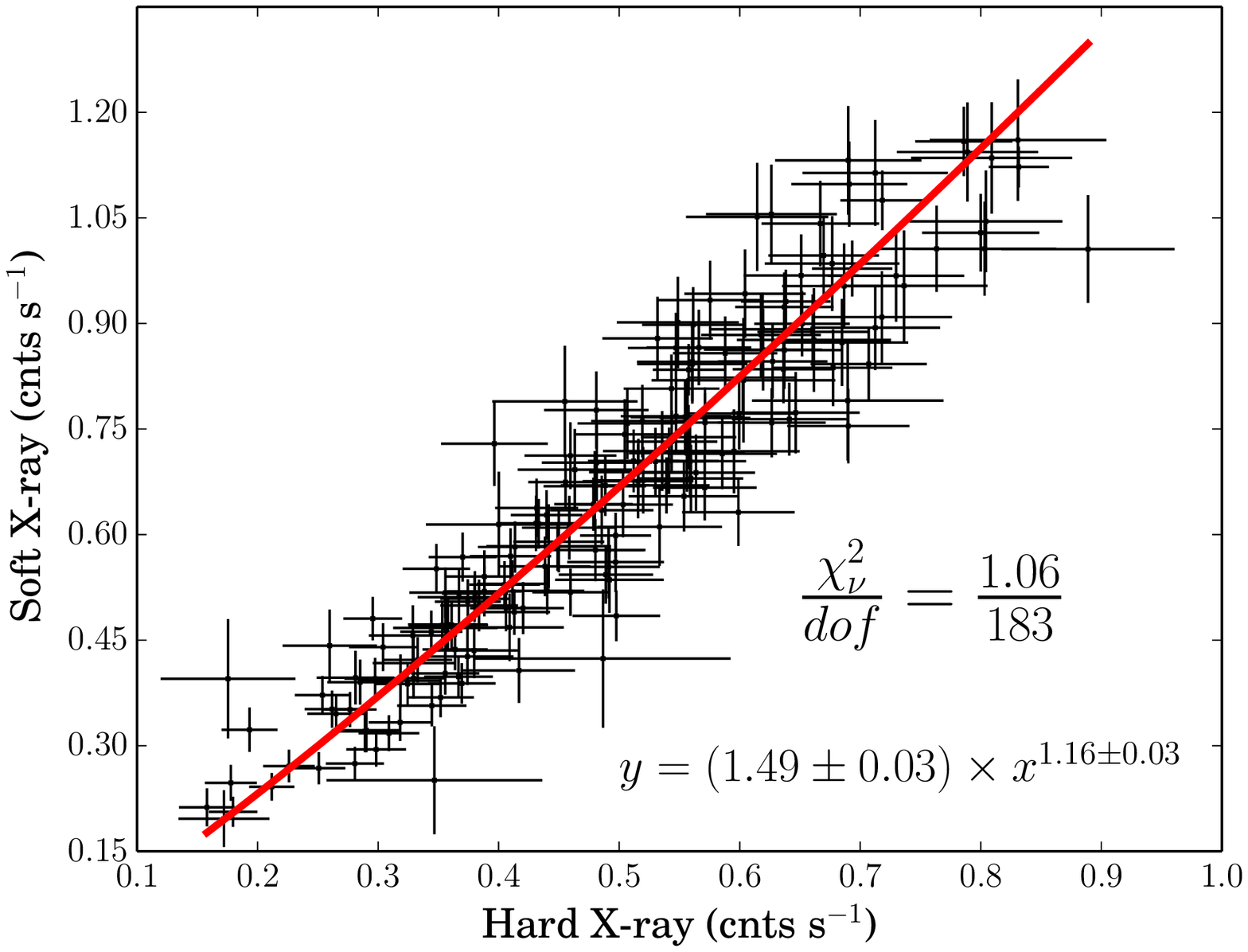}
\includegraphics[scale=0.42]{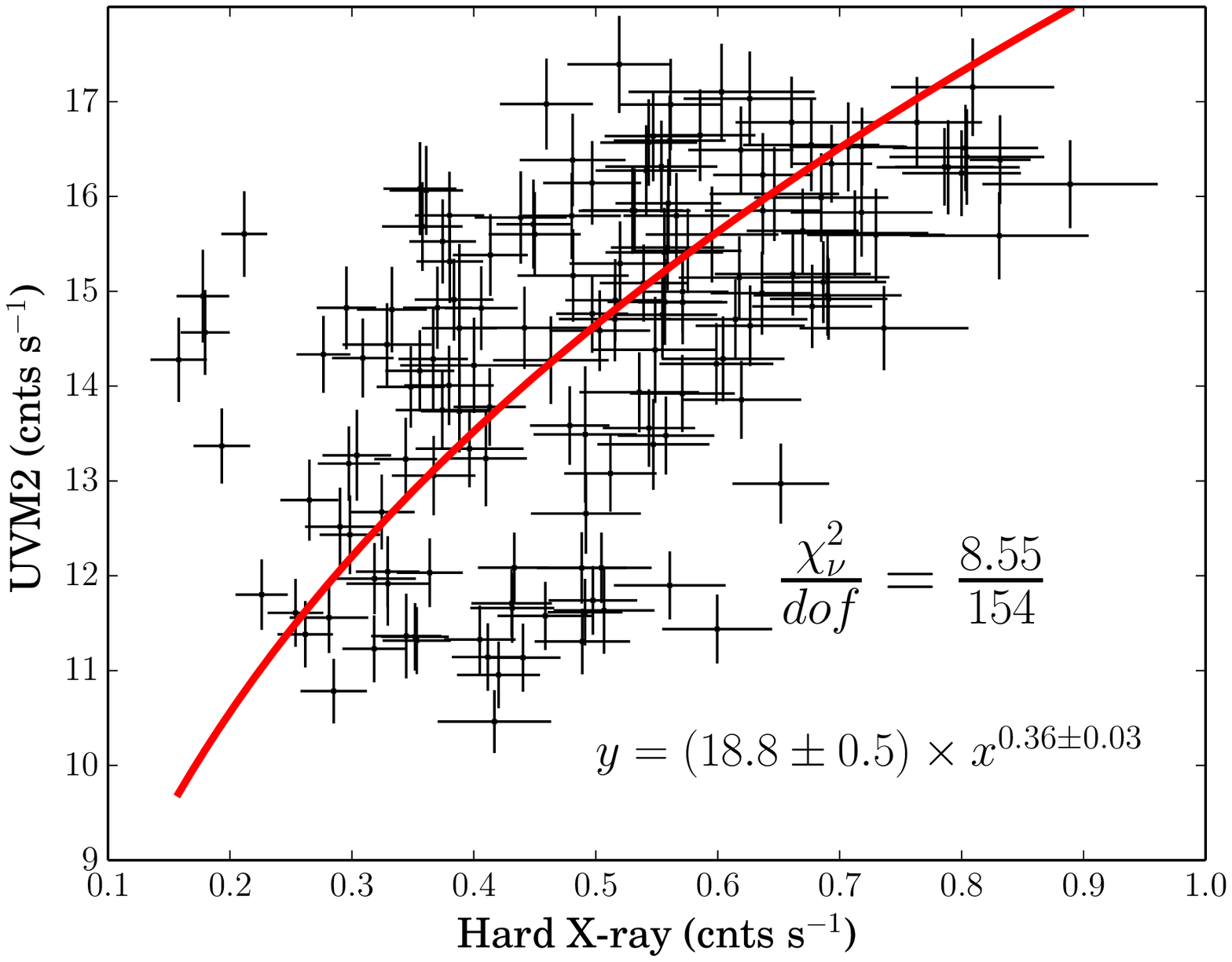}
\caption{Count--count plots for the soft X--ray and hard X--ray (left panel), and the UVW2 and hard X--ray (right panel) bands are shown for the modeling with a power law $y=(1.49\pm0.03) \times x^{1.16\pm0.03}$ and $y=(18.8\pm0.5) \times x^{0.36\pm0.03}$, respectively.}
\label{absCheck} 
\end{figure*}

\subsection{ Spectral analysis}
The \swift{}/XRT, which covers 0.3-10 keV energy range, is appropriate to study 
the presence of several spectral components such as primary power law continuum, 
soft X-ray excess and Fe-K$\alpha$ emission line along with the signature of 
absorption components in the soft band. Presence of soft X-ray excess and warm
absorber have been detected in the galaxy NGC~4593 \citep{2016MNRAS.463..382U,2013MNRAS.435.3028E}. We used phenomenological models 
such as blackbody ({\it bbody} in {\tt XSPEC}) for the soft X-ray excess and 
power law ({\it powerlaw in {\tt XSPEC}}) continuum to describe the X-ray 
broad-band spectra from all the observations used in the present work. We 
multiplied photo-absorption model ({\it phabs} in {\tt XSPEC}) to modify the 
soft X-ray band due to the Galactic absorption. Thus, we used 
{\it phabs(bbody+powerlaw)} model to fit spectra from all the observations 
except those which were highly piled-up. We found that this composite model 
is sufficient to describe the broad-band spectrum of NGC~4593 without any 
requirement of warm absorber or Fe-K$\alpha$ emission line. A representative
spectrum of the galaxy (Obs. ID-00092353119) is shown in Fig.~\ref{bestfit} 
along with the best-fit model components and residuals obtained from our 
fitting. The distribution of reduced $\chi^{2}$ values obtained from the 
spectral fitting of all the observations used in present work is displayed 
in upper panel of Fig.~\ref{dist}. We obtained similar correlation between 
the fluxes of soft X-ray excess and power law continuum as we found in the 
count--count plot as shown in Fig.~\ref{corr}(a). We could not constrained 
the power-law photon index for all the observations due to low exposure. 
The lower panel of Fig.~\ref{dist} shows the variation of the power-law 
photon index $\Gamma$ over the observation campaign and found to be 
consistent as reported in \citet{2016MNRAS.463..382U}.

\section{Discussion}

 We analyzed the variability observed in the UV/optical and X-ray bands of a Seyfert~1 galaxy NGC~4593 using archival intensive \swift{} campaign during July 13 -- August 5, 2016. We applied various tools to investigate variability seen the UV/optical and X-ray light curves. Linear correlation between the $1.5-10\kev$ and one of the low energy i.e.,  the UV/optical bands including soft X-ray band was used in our work. We studied the count-count correlation with positive offset (C3PO) method to distinguish less variable component compared to X-ray emission. Cross-correlation analysis was done using {\tt ZDCF} and {\tt JAVELIN} and derived the lags for the soft bands compared to hard X-ray band. We summarize our main results which are as follows: 

\begin{enumerate}[(i)]

\item NGC~4593 exhibited strong variability in all bands on short and long timescales. 

\item The linear fit to the correlation between the hardness ratio and the hard X-ray count rate (see Fig.~\ref{sw_obs1}) suggests that the source retains its nature of being softer when brighter as found by \citet{2016MNRAS.463..382U}. 

\item The negative offset from C3PO technique favors a weakly absorbed soft X-ray band and is consistent as found by \citet{2016MNRAS.463..382U}. The positive offset in linear fit for the UV/optical emission also suggests the presence of weakly variable component in the UV/optical bands possibly the disc emission (see Fig.~\ref{corr}).

\item Linear correlation  coefficient ($\rho=0.29-0.94$) and low p value ($10^{-4}-10^{-84}$) infer that the observed light curves in the UV/optical bands are strongly correlated with the hard X-ray light curve. 

\item The UV/optical emission follow the hard X-ray emission by $\sim 0.4-1.5{\rm~days}$ as estimated by the cross-correlation analysis. The lag estimated for each pair of light curves by both {\tt ZDCF} and {\tt JAVELIN} are found to be comparable. 

\item The observed lag increases with wavelength despite large errors. The power law model with $4/3$rd power of wavelength describes well to the observed lag spectrum (lag vs. wavelength).

\item The observed lags are found to be larger than the derived lags from Shakura-Sunyaev disc theory after adding both the thermal and X-ray heating (see Fig.~\ref{lag_spec}).
\end{enumerate}

The observed light curves in the X-ray as well as UV/optical bands appear highly variable on both long and short timescales. The X-ray emission varies by a factor of $\sim5$ and the UV/optical emission vary by $\sim$80--20\%~over about a month timescale. The X-ray emission fluctuates rapidly by a factor of $\sim$ 2 while the UV/optical emission vary slowly ($\sim$27--18\%). Such observed variations may be associated with various phenomena such as extinction/absorption along the line of sight, the changes in the accretion flow, Comptonization of seed photons in the disc and X-ray reprocessing in the accretion disc. The hardness ratio (right bottom panel of Fig~\ref{uvobs1}) and positive offset in the linear fit (Fig.~\ref{corr}) imply that there is no significant effect of absorption/extinction 
on the emission on a month timescale. The observed variation in hardness ratio on hours 
time scale (right bottom panel of Fig~\ref{uvobs1}) indicates the presence of variable absorption in the galaxy. Similarly, from the count-count correlations (Fig~\ref{corr}), 
scatter of data points from the linear fit in UV/optical bands also suggests the 
possible presence of absorption.

A careful investigation of the scatters suggests that the distribution of data points 
in count-count correlation plots (Fig~\ref{corr}) follows a non-linear trend (except 
soft X-ray band and V band). This indicates that the variability observed in above bands 
may be due to absorption along the line of sight. For example, the negative offset in 
soft X-ray band can be verified by assuming that both soft and hard X-ray bands are 
affected by absorption. If that is the case, the simple power law continuum model can
be modified by incorporating absorption component. This implies, in soft X-ray band, 
the observed count rate $S(t)$ at energy $E_S$ can be expressed as:
\begin{equation}
S(t) = \left(\{\prod_{i=1}^{N} \exp\left[-N_{\rm H,i}(t)\sigma(E_S)\right]\right\} C E_S^{-\Gamma},
\end{equation} 
\noindent
where $N$ is the number of obscuring clouds along the line of sight with variable equivalent hydrogen column density $N_{H,i}(t)$. We also assume that the shape of the X-ray continuum remains constant. $\sigma(E_S)$ and $C$ are the photo-electric cross-section in soft energy band and normalization constant, respectively. The above equation can be rewritten as, 

\begin{equation}
S(t) =\exp\left\{\left[-\sum_{i=1}^{N}N_{\rm H,i}(t)\right]\sigma(E_S)\right\} C E_S^{-\Gamma},
\label{ynh}
\end{equation}
\noindent
and similarly the same will be the case for the count rate in hard X-ray band at energy $E_H$,
\begin{equation}
H(t) = \exp\left\{\left[-\sum_{i=1}^{N} N_{\rm H,i}(t)\right]\sigma(E_H)\right\} C E_H^{-\Gamma}.
\label{xnh}
\end{equation}
\noindent

We can rewrite above equations by eliminating $\left[-\sum N_{\rm H,i}(t)\right]$ to get the following expression between the count rates in the two bands,
\begin{equation}
S = A H^{\beta},
\label{xynh}
\end{equation}
\noindent
where $A$ is another normalization constant and slope $\beta$ is described as $\beta=\sigma(E_S)/\sigma(E_H)$. Eq. \ref{xynh} predicts a  non linear relation between hard and soft bands. Even if number of clouds $N$ changes with time, Eq.\,\ref{xynh} should be valid.

We used Eq. \ref{xynh} to test the absorption (irrespective of either warm absorption or neutral) in the soft and hard X-ray bands. We modeled the count-count correlation assuming 
soft X-ray counts as the function of hard X-ray counts and obtained a marginal improvement 
compared to the earlier linear fit. It is interesting to note that the slope of the power 
law fit ($\sim$1.16, as shown in left panel of Fig.~\ref{absCheck}) is consistent with the
photo-absorption cross-section dependency on energy ($\sigma\propto E^{-3}$). This suggests
that the soft X-ray band is mildly affected from absorption, supporting the negative offset
finding in count-count correlation study.

It is interesting to check similar effect in the UV/optical bands where we noticed non-linear
trends. We modeled the count-count correlation between the UVM2 band (as the soft band) and 
the hard X-ray band and presented in the right panel of Fig.~\ref{absCheck}. We found significant improvement ($\frac{\chi^{2}}{dof}=$1316.7/154) in statistics compared to 
the linear fit ($\frac{\chi^{2}}{dof}=$1336.4./154). Interestingly, the slope $\beta\sim0.36$ 
was found to be lesser than one. This implies that the photo-absorption cross-section 
is higher in the hard X-ray band compared to the UVM2 band. Similar fits were also found 
in other UV/optical bands. Such findings are contrary to the theory of photo-absorption 
where photo-absorption cross section $\sigma\propto E^{-3}$. This suggests that the 
scatters present in the UV/optical bands are not caused due to the line of sight 
absorption. These scatters may be due to diffused emission from broad-line regions.         

On the other hand, strong correlation between the soft bands and the hard X-ray band infers that these emission are related to either changes in the accretion flow, Comptonization phenomena or X-ray reprocessing. A month timescale is too short to see any
significant variation caused by the changes in the accretion flow as the expected timescale 
of fluctuations in the accretion flow is $\sim10^3$ years or more \citep{2007MNRAS.375.1479S,2017arXiv171001944P}. The observed short timescale variation can be caused by either inverse Comptonization or the X-ray reprocessing phenomenon in the disc. In the inverse Comptonization process, the variations in the UV/optical emission should lead the variations in the X-ray emission on the light-crossing timescale. Similarly, the X-ray reprocessing causes the short timescale in such a way that the changes in the UV/optical emission lags the fluctuations observed in the X-ray emission.

Cross-correlation analysis using {\it ZDCF} shows that the observed changes in the UV/optical bands lags the observed variations in the X-ray bands on short timescale of few days ($\sim0.4-1.5$ days). Thus the inverse Comptonization process may not be working between the various bands. This implies that X-ray reprocessing is playing a major role to drive the observed variations on short timescale of few days ($\sim0.4-1.5$ days). Such lags are very important and interesting to study various regions of the accretion disc and its coupling with the X-ray corona. The lags provide the distribution of regions on the accretion disc which emit at various wavelengths. The observed lag as a function of wavelength is well described by 4/3rd power of the wavelength. This power law description of the lag spectrum is consistent with the standard accretion disc model (see blue curve in Fig.~\ref{lag_spec}). However, the estimated lags from the Shakura-Sunyaev accretion disc theory under lamp-post geometry assumption appear smaller than the observed lags (see green dashed line in Fig.~\ref{lag_spec}). For example, UVW1, U and B bands show clear longer lags than that of the predicted values from standard disc model while for Soft X-ray, UVW2, UVM2 and V bands, both the values marginally agree. Though, the estimated lags are determined by considering wavelength calculated from Wein's law, such assumption provides overestimation of lags by not including the geometrical effects (e.g., flux weighted mean radius, inclination) as presented in \citet{2014MNRAS.444.1469M}. By adding geometrical effects, the predicted lags would be further shortened.

From the observed time lag, the size of the disc is found to be larger compared to the 
standard disc prediction. There is a chance that the standard disc prediction might be underestimated due to the assumptions such as the disc is (i) optically thick and behaves 
like a blackbody and (ii) geometrically thin i.e. height to radius ratio is very small.
In addition to above assumptions, in lamp-post geometry, the isotropically emitting corona 
is in compact form and lies on the spin axis of black hole. However, actual geometry 
of the disc and corona may be different. The disc may be a flared disc (e.g., \citealt{1976ApJ...208..534C}) and the corona may be in extended form (e.g., \citealt{2013MNRAS.435.1287P,2001MNRAS.327..799K,2012MNRAS.419.2369I}). Figure~1 of \citet{2017MNRAS.470.3591G} shows that if the extended corona illuminates the disc, such 
illumination can make the disc a flared disc. In that case, the wavelength dependent 
lag follows $\lambda^{7/3}$ profile rather than the $\lambda^{4/3}$ profile. This 
implies the observed lag may be due to enhanced flux from the outer disc region. 
\citet{2017MNRAS.470.3591G} reported that the observed variation in UV/optical 
emission from NGC~5548 is not consistent with the reprocessing of hard X-rays 
rather it is due to the reprocessing of FUV emission from the inner region. Similar 
results were also reported in NGC~4151 \citep{2017ApJ...840...41E}. 
\citet{2017ApJ...846...55M} did not find any correlation between the continuum 
flux and emission line flux from {\it Chandra} and {\it Swift} observations of 
NGC~5548. They suggested that such anomaly is possibly caused due to the presence 
of warm Comptonizing medium in the inner disc which is consistent with the scenario 
given by \citet{2017MNRAS.470.3591G}. We also tried to investigate the possibility 
of second reprocessor by excluding the X-ray lags with respect to the UVW2 band. 
However, our results are not consistent with the possibility of second reprocessor 
(see right panel Fig.~\ref{lag_spec}).

The findings presented in this paper are interesting in a sense that our knowledge 
of real accretion theory is not complete and the actual accretion disc appears larger 
in size than that predicted by standard disc model. In addition to the 
lag-wavelength profile of $\lambda^{4/3}$, we found larger lag in the U band. This 
suggests an additional component of emission contributing in this band. This emission 
component may be diffused emission from broad line region. The nature of real accretion 
disc could be very complex and may provide longer lag due to its inhomogeneity of the outer surface and warped nature \citep{2011ApJ...727L..24D}. The micro-lensing results also 
suggest that the observed emission requires larger region of the outer accretion disc 
over standard accretion disc \citep{2010ApJ...712.1129M}.  The power-law 
index was fixed at 4/3 in our analysis due to the fact that this parameter is not well constrained given the available data. In addition, the large errors in the lag estimation 
can be improved by having more observations of the source. However, better estimation of
lag alone is not sufficient to constrain the physical size of the accretion disc in AGN.
Along with the estimated lag, results from the effect of general relativity, geometrical effects, complex absorption near the accretion disc and disc-corona interaction together 
can provide reliable estimation of the disc size.
           
\section{Acknowledgment}
We sincerely thank the reviewer for his/her useful comments and suggestions which 
improved the paper significantly. The 
High Energy Astrophysics Science Archive Research Center (HEASARC) of the NASA/GSFC
Astrophysics Science Division and Smithsonian Astrophysical Observatory's High Energy
Astrophysics Division were used for their available softwares and online webtools. 
This research has made use of the archival observations of \swift{} observatory and
XRT Data Analysis Software ({\tt XRTDAS}) developed under the responsibility of the 
ASI Science Data Center (ASDC), Italy.

\newcommand{\pasp}{PASP} \def\apj{ApJ} \def\mnras{MNRAS} \def\astl{AstL}
\def\aap{A\&A} \def\apjl{ApJ} \def\aaps{A\&AS}\def\aj{aj} \def\physrep{PhR}
\def\pre{PhRvE} \def\apjs{ApJS} \def\pasa{PASA} \def\pasj{PASJ}
\def\nat{Nat} \def\ssr{SSRv} \def\aapr{AAPR} \def\araa{ARAA} \def\procspie{SPIE}

\bibliographystyle{mn2e} 
\bibliography{refs}

\end{document}